\documentclass[preprint,11pt,3p,authoryear]{elsarticle}

\usepackage{natbib}
\usepackage{svg}
\usepackage{url}
\usepackage{rotating} 
\usepackage{amssymb}
\usepackage{amsmath}
\usepackage{gensymb}
\usepackage{booktabs}
\usepackage{multirow}
\usepackage{threeparttable}
\usepackage{graphicx}
\usepackage{caption}
\usepackage{array}
\usepackage{subcaption}
\usepackage{cleveref}
\usepackage{pdflscape}
\usepackage{chngcntr}

\usepackage{amsthm}
\usepackage{geometry}
\usepackage{tabularray}
\usepackage{makecell}
\usepackage{setspace}
\usepackage{comment}
\usepackage{tabularx}      
\usepackage{adjustbox}     
\usepackage{lineno}

\begin{document}
\doublespacing
\begin{frontmatter}

\title{Differentiable Graph Neural Network Simulator for the Back-Analysis of Post-Liquefaction Residual Strength from Flow Failure Runout}

\author[label1]{Yongjin Choi}
\author[label1]{Jorge Macedo}

\affiliation[label1]{organization={School of Civil and Environmental Engineering, Atlanta, Georgia Institute of Technology},
            city={Atlanta},
            postcode={30332}, 
            state={GA},
            country={USA}}

\begin{abstract}
This study introduces Differentiable Graph Neural Network Simulators (Diff-GNS) as a physics-informed and automated framework for estimating post-liquefaction residual strengths ($S_r$). Traditional approaches to estimate $S_r$ rely on simplified physics, manual iterations, and assumptions about runout development. Diff-GNS overcomes these limitations by integrating a Graph Neural Network Simulator (GNS) that simulates granular flows, with gradient-based optimization through automatic differentiation. GNS accelerates forward runout simulations that are otherwise computationally intensive with conventional numerical methods, while gradient-based optimization automates the inversion to back-calculate $S_r$. The GNS is trained on simulations with the material point method on geometries informed by case-history runout failures, enabling focused learning of realistic runout mechanisms and the ability to simulate slopes across small and large scales. The Diff-GNS framework is validated using two well-documented liquefaction-induced flow failure case histories: the Lower San Fernando dam and La Marquesa dam. In the two cases, the inferred $S_r$ agrees closely with published estimates and reproduces physically consistent runout behaviors. The framework also has the ability to jointly infer multiple interacting parameters, extending beyond single-parameter back-analyses. By embedding the physics of runout processes, minimizing manual intervention, and accelerating the inversion process to estimate $S_r$, Diff-GNS provides an efficient, reproducible, and physically grounded approach for geotechnical analysis of liquefaction-induced flow failures.
\end{abstract}


\end{frontmatter}


\section{Introduction}

Over the past decades, methods for estimating the post-liquefaction residual shear strength ($S_r$) and evaluating post-failure runout in slope systems have evolved. Initial reliance on laboratory testing to estimate $S_r$ faced challenges in replicating in-situ field conditions, particularly the effects of void redistribution and sample disturbance \citep{poulos1985liquefaction,castro1992steady}. These difficulties steered the focus toward the back-analysis of full-scale field case histories. Back analyses typically use field observation data, i.e., failure geometry and runout characteristics, to infer the operational shear strengths of the soil mass that can explain the observed failure runout. Early back-analysis methods \citep{seed1987design, seed1990spt} were based on simple limit equilibrium principles applied to the static geometries of the slope before and after failure. However, static analyses neglect the kinetic energy and momentum carried by the sliding mass during failure, known as momentum effects \citep{olson2002liquefied}. 

Previous efforts to consider momentum effects in back-analysis include the kinetics method \citep{olson2001liquefaction_kinetics,olson2002liquefied}, the zero inertial factor (ZIF) method \citep{kramer2015empirical_zif}, and the incremental momentum method (IMM) \citep{weber2015engineering_imm}. These methods have been used to evaluate well-known case histories; however, each method faces limitations. The kinetics method, for example, simplifies the failure mass to a rigid block with a pre-assumed sliding path. The ZIF method relies on an intermediate sliding mass at an inferred instance of zero acceleration, with several assumptions for its definition. The IMM involves manually constructing a sequence of kinematically feasible cross-sections, making the analysis subject to a range of plausible interpretations and time-consuming. 

Addressing the highlighted shortcomings calls for a paradigm shift to assess runout flow failures and $S_r$, from manual, interpretive techniques that can involve significant judgment, toward a more automated, physics-consistent, and computationally efficient approach. Numerical techniques capable of simulating large deformations, such as the Material Point Method (MPM), can offer a more physically rigorous foundation \citep{macedo2024mpm-granular, de2020mpm}. However, they are not computationally efficient, runtimes can vary from hours to days \citep{abram2022mpm_insitu,ceccato2024simulating_mpm}, limiting their practical application. 
This computational burden becomes especially prohibitive in back analyses that seek to estimate $S_r$, because they would require repeated forward simulations to iteratively refine strength parameters. 

Addressing the limitations of simplified methods and the high computational cost of numerical approaches such as MPM, this study introduces a computationally efficient framework for back-calculating $S_r$ while maintaining a physics-informed runout progression during failure, leveraging the Graph Neural Network simulator (GNS) concept developed by DeepMind \citep{sanchez2020learning}. The framework couples a trained GNS in this study with gradient-based optimization via reverse-mode automatic differentiation (AD) \citep{baydin2018ad}, yielding a Differentiable Graph Neural Network Simulator (Diff-GNS) capable of inferring material parameters efficiently. Previous GNS applications \citep{choi2024graph, choi2023three, zhao2025physical} have demonstrated the GNS potential as an efficient surrogate for granular flow simulations, but remained largely proof-of-concept, limited to small domains (1-2 m), idealized cube-shaped masses, and homogeneous materials characterized only by friction angle. The only study we know that extended GNS to realistic slope systems (\cite{choi2025differentiable_multilayers}) provided encouraging results but relied on training with randomly generated geometries, not directly related to slope case history failures. In contrast, this study designs the training datasets informed by case histories and develops distinct models for small and large-scale slopes, enabling representation of systems spanning tens to hundreds of meters, a capability not included in earlier efforts. These advances move GNS from conceptual validation toward a practical tool for estimating $S_r$ and simulating runout in slope systems. We evaluate the Diff-GNS's performance in back-analyzing $S_r$ and simulating runout dynamics, considering two liquefaction flow failures: the Lower San Fernando dam and the La Marquesa dam. To our knowledge, this is the first assessment of Diff-GNN for real-world runout case histories. 

The manuscript is organized as follows: \Cref{sec:existing_methods} reviews traditional $S_r$ estimation methods and their limitations. \Cref{sec:diff-gns} presents the theoretical basis of the Diff-GNS framework and its advantages. \Cref{sec:training} describes the training of the GNS implemented in this study. \Cref{sec:results} evaluates the trained GNS's performance in estimating $S_r$ and simulating realistic runout for selected case histories.

\section{Existing methods to back-calculate residual strength}\label{sec:existing_methods}

\begin{figure}[!htbp]
    \centering
    \includegraphics[width=1.0\textwidth]{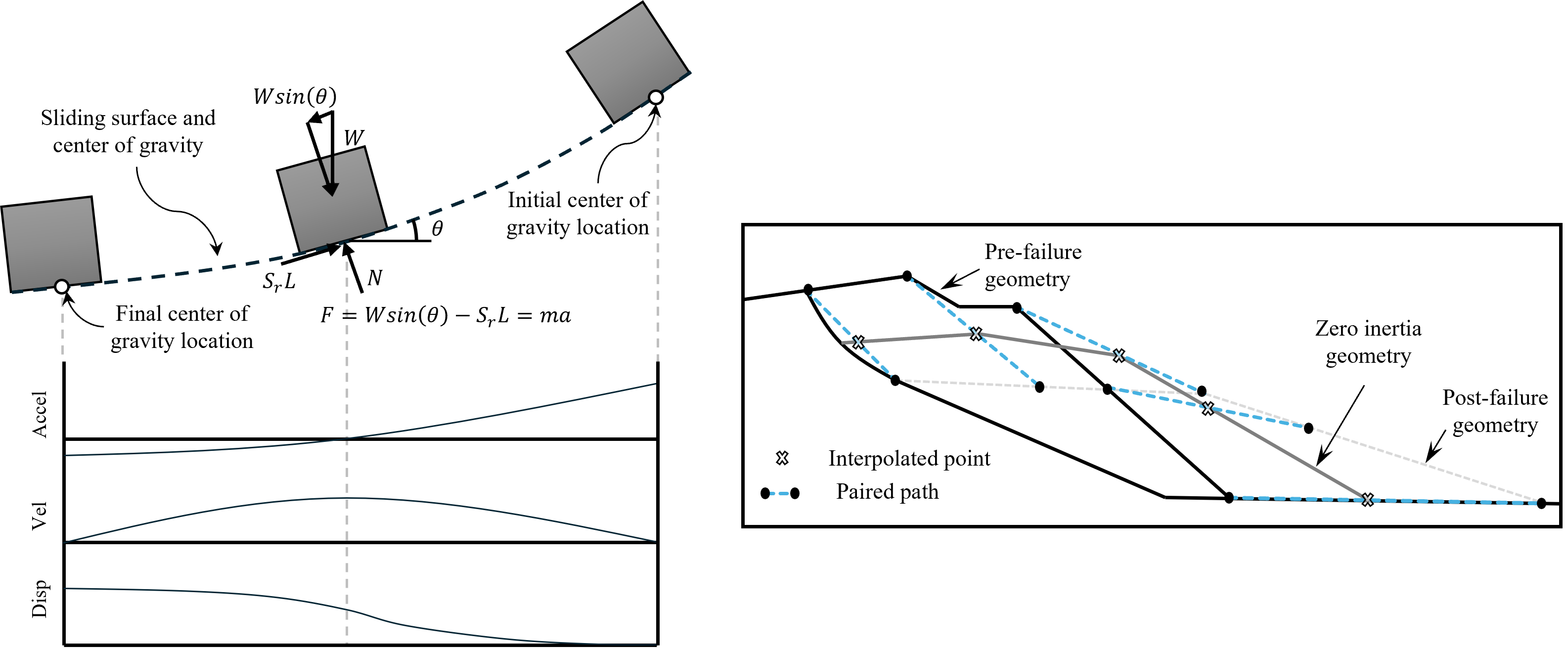}
    \caption{Conceptual illustration of (a) the Kinetics method (after \cite{olson2001liquefaction_kinetics}) and (b) the Zero Inertial Factor (ZIF) method (after \cite{kramer2015empirical_zif}) for back-analysis of post-liquefaction residual strength $S_r$.}
    \label{fig:kinetics_and_zif_method}
\end{figure}

\paragraph{\textbf{Kinetics Method}} The kinetics method \citep{olson2001liquefaction_kinetics, olson2002liquefied}, illustrated in \Cref{fig:kinetics_and_zif_method}a, applies dynamic equilibrium ($\sum F = m \cdot a$) to back-calculate $S_r$, explicitly accounting for momentum effects. The failed mass is idealized as a rigid body translating downslope, with its center-of-gravity (CG) path approximated by a third-order polynomial. The slope of this curve defines the driving force ($W\sin\theta$), while resistance equals $S_r L$. Incremental time-stepping analysis updates acceleration, velocity, and displacement to track the rigid body's motion. $S_r$ is iteratively adjusted until the final computed CG position matches field observations. The method is sensitive to the assumed CG path and relies on rigid-body idealized motion.

\paragraph{\textbf{Zero Inertial Factor (ZIF) Method}} The ZIF method \citep{kramer2015empirical_zif}, illustrated in \Cref{fig:kinetics_and_zif_method}b, provides a static approach to implicitly account for the momentum effect by identifying the motion stage where acceleration is zero and forces balance. ZIF is defined as the fraction of total displacement at this stage, typically assumed to be 40-50\%, or estimated from the kinetics method. 
To build the ZIF geometry (gray line in \Cref{fig:kinetics_and_zif_method}b), points from the pre-failure cross-section (black dots) are shifted toward the post-failure geometry according to the ZIF value, and the displaced points ($\boldsymbol{\times}$ markers) define ZIF. Static limit-equilibrium analysis is then performed on this geometry to back-calculate $S_r$ with a safety factor of 1.0. The method avoids dynamic modeling but depends heavily on engineering judgment to define the ZIF geometry, which may require iteration and limit reproducibility.

\paragraph{\textbf{Incremental Momentum Method (IMM)}} 

IMM \citep{weber2015engineering_imm} addresses limitations of the kinetics and ZIF methods by reconstructing a sequence of intermediate geometries that progressively track the slope from its initial to final state. At each stage, driving and resisting forces are calculated, and Newton’s second law is applied incrementally to update acceleration, velocity, and displacement until the modeled runout matches observations. This stepwise approach captures evolving geometries, accommodates progressive failure, and avoids curve-fitting assumptions such as polynomial CG paths. However, it is labor-intensive and relies heavily on engineering judgment to define intermediate physically plausible stages, making results dependent on the analyst.

\section{Differentiable graph neural network simulator-based framework}\label{sec:diff-gns}

\subsection{Graph Neural Network Simulator (GNS) for Granular Flow}

\begin{figure}[!htbp]
    \centering
    \includegraphics[width=0.8\textwidth]{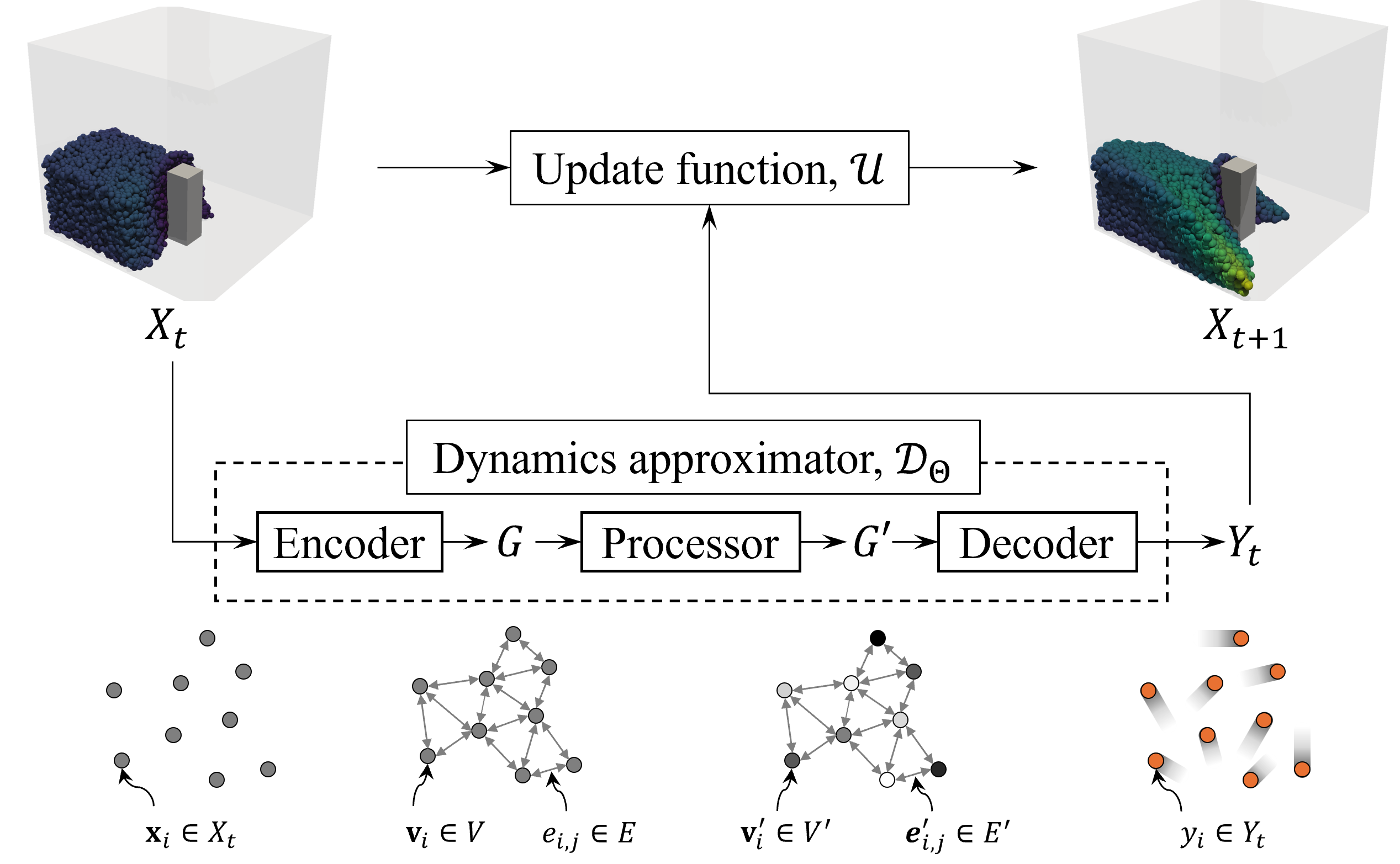}
    \caption{Schematic of the Graph Neural Network Simulator (GNS) architecture.}
    \label{fig:gns}
\end{figure}

GNS can be used as a generalizable surrogate for conventional numerical solvers that simulate granular flows. Below, we briefly introduce the GNS architecture using \Cref{fig:gns}. For further details, see the original works from \cite{sanchez2020learning} and \cite{choi2024graph}.

Consider a physical state of a granular flow at time $t$, $X_t = \{ \mathbf{x}_i^t \}_{i\in N}$, where each $\mathbf{x}_i^t = [\mathbf{p}_i^{t-k:t}, \mathbf{b}_i^t, \mathbf{f}, \mathbf{w}]$ is the feature of the material point $i$ in $N$ total material points. Here, $ \mathbf{p}_i^{t-k:t} $ denotes the material point positions from time $ t-k $ to $ t $, $ \mathbf{b}_i^t $ is boundary information, $ \mathbf{f} $ is a categorical identifier for the material point type (e.g., fixed or kinematic), and $ \mathbf{w} $ contains material properties i.e., $ \mathbf{w} = [\phi_i, c_i] $ for the Mohr-Coulomb materials used in this study.

The GNS represents $X_t$ as a graph $G = (V, E)$, where vertices $V = \{ \mathbf{v}_i \}_{i\in N}$ correspond to material points and edges $E = \{ \mathbf{e}_{i,j} \}_{i\in N, \ j\in\mathcal{N}(i)}$ describe the interactions between the material point $i$ and its neighbor $j$, where $\mathcal{N}(i)$ represents the set of vertices connected to the vertex $i$. The GNS updates $G$ to $G'=(V', E')$ based on the learned local interaction law modeled through a message passing graph neural network (GNN) \citep{battaglia2018inductive}, and returns the next state $X_{t+1}$. The GNS consists of a learned dynamics approximator $\mathcal{D}_\Theta$ and the update function $ \mathcal{U} $. $\mathcal{D}_\Theta$ adopts a encoder-processor-decoder structure. First, the encoder takes $X_t$ and embeds it into the latent graph $G$. Next, the processor performs message passing with GNN, which propagates the vertices' information to their neighbors along edges, returning the updated graph $G'$. This message passing models the momentum or energy exchange between material points. Finally, the decoder extracts the dynamics information $Y_t = \{ \mathbf{y}_i^t \}_{i\in N}$ of the material points from the updated graph $G'$. The update function $ \mathcal{U} $ then advances the current state $X_t$ using the predicted dynamics $ Y_t$ (i.e., $X_{t+1} = \mathcal{U}(X_t, Y_t)$).
GNS recursively predicts the next states until a desired timestep $k$, called rollout:

\begin{equation}\label{eq:gns}
    X_{t+1} = \text{GNS}(X_t), \quad X_0 \rightarrow X_1 \rightarrow \cdots \rightarrow X_k
\end{equation}

All modules in $\mathcal{D}_\Theta$---encoder, processor, and decoder---are implemented as multilayer perceptrons (MLPs) with two hidden layers of 128 units, comprising the learnable parameter set $\Theta$. The training objective \cref{eq:gns_objective} is to minimize the mean squared error (MSE) between the predicted dynamics $\mathbf{y}_i^t \in Y_t$ and the ground truth accelerations of material points $\mathbf{a}_i^t \in A_t$ by optimizing $\Theta$. 

\begin{equation}\label{eq:gns_objective}
    \mathcal{J}(\Theta) = \frac{1}{N} \sum_{i=1}^{N} ||\mathbf{y}_i^t - \mathbf{a}_i^t||^2
\end{equation}

In this study, we generate the training data using MPM, which will be explained in \Cref{sec:training}.

Once trained, GNS serves as a surrogate for MPM. GNS can simulate a broad range of granular flow configurations not encountered during training, achieving orders-of-magnitude speedup compared to MPM, while retaining physical fidelity by learning the underlying interaction laws that govern granular flow dynamics \citep{choi2024graph,zhao2025physical}.

\subsection{Inverse analysis framework with differentiable GNS}

\begin{figure}[!htbp]
    \centering
    \includegraphics[width=0.95\textwidth]{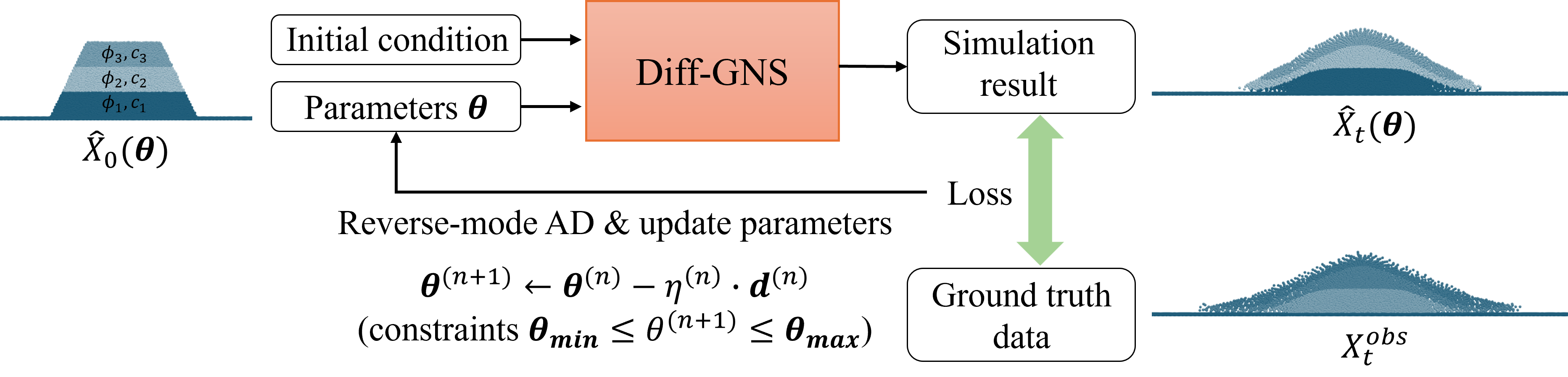}
    \caption{Differentiable graph neural network simulator (Diff-GNS)-based framework for solving inverse problems in granular flows (after \cite{choi2025differentiable_multilayers}). Given the initial condition and parameters $\boldsymbol{\theta}$ for a slope system at time $0$, $ \hat{\mathbf{X}}_0(\boldsymbol{\theta}) $, Diff-GNS runs the forward simulation and generates $ \hat{\mathbf{X}}_t(\boldsymbol{\theta}) $, granular flow state at time $t$.  It then evaluates the loss compared to $X_t^{\text{obs}}$, the observed state. $\boldsymbol{\theta}$ is updated using gradient-based optimization with optional given constraints.}
    \label{fig:diff-gns}
\end{figure}

The inversion is based on a differentiable GNS that uses gradient-based optimization through automatic differentiation (\Cref{fig:diff-gns}). A brief description follows; further details are provided in \cite{choi2024inverse}, \cite{choi2025differentiable_multilayers}, and \cite{baydin2018ad}.

Given a set of observed granular flow states $ \{X_t^{\text{obs}} \}_{t \in \mathcal{T}} $ at a subset of time indices (e.g., $ \mathcal{T} \subseteq \{0, 1, \dots, T\} $), the goal is to estimate the optimal parameters $ \boldsymbol{\theta}^* $ that minimize the discrepancy between the observed states and the corresponding forward simulation results from GNS:

\begin{equation}\label{eq:inverse_loss}
    \boldsymbol{\theta}^* = \arg\min_{\boldsymbol{\theta}} \; \mathcal{L}_{\boldsymbol{\theta}} \left( \{X_t^{\text{obs}} \}_{t \in \mathcal{T}},\; \{ \hat{X}_t(\boldsymbol{\theta}) \}_{t \in \mathcal{T}} \right)
\end{equation}

Here, $ \hat{X}_t(\boldsymbol{\theta}) $ denotes the GNS-simulated granular flow state at time $ t $ under parameters $ \boldsymbol{\theta} $, and $ \mathcal{L}_{\boldsymbol{\theta}}(\cdot, \cdot) $ is a task-specific loss function measuring the mismatch between observed and simulated states. In this study, 
$ \boldsymbol{\theta} $ is a set of strength parameters, such as $ S_r $ or friction angle $\phi$; $ \mathcal{L} $ is the runout distance. The time index set $ \mathcal{T} $ refers to the condition with information; in the considered case histories, only the final post-failure geometry is available, in which case $ \mathcal{T} = \{T\} $.

Diff-GNS solves \Cref{eq:inverse_loss} using gradient-based optimization (\Cref{eq:gd_general}). It iteratively improves $\boldsymbol{\theta}$ using the gradient of the loss $\mathcal{L}_{\boldsymbol{\theta}}$ with respect to $\boldsymbol{\theta}$, $\nabla_{\boldsymbol{\theta}} \mathcal{L}_{\boldsymbol{\theta}}$ (\Cref{eq:grad_loss}).

\begin{equation}\label{eq:gd_general}
    \boldsymbol{\theta}^{(n+1)} \leftarrow \boldsymbol{\theta}^{(n)} - \eta^{(n)}\cdot \mathbf{d}^{(n)}
\end{equation}

\begin{equation}\label{eq:grad_loss}
    \nabla_{\boldsymbol{\theta}} \mathcal{L}_{\boldsymbol{\theta}} = \frac{\partial \mathcal{L}_{\boldsymbol{\theta}}\left(\{X_t^{\text{obs}}\}, \{\hat{X}_t(\boldsymbol{\theta})\}\right)}{\partial \boldsymbol{\theta}}
\end{equation}

Here, $n$ is the optimization step, $\eta^{(n)}$ is the learning rate, and $\mathbf{d}^{(n)}$ is the update direction, which is a function of  $\nabla_{\boldsymbol{\theta}} \mathcal{L}_{\boldsymbol{\theta}}$. Because GNS is implemented as a neural network, it is fully differentiable and allows efficient computation of $\nabla_{\boldsymbol{\theta}} \mathcal{L}_{\boldsymbol{\theta}}$ using AD.

There is often prior knowledge of feasible parameter ranges. 
Embedding these bounds in the optimization both reduces the search space and avoids nonphysical solutions. We adopt the L-BFGS-B algorithm \citep{byrd1995lbfgs-b}, a limited-memory quasi-Newton method that supports box constraints of the form $\boldsymbol{\theta}_{min} \leq \boldsymbol{\theta} \leq \boldsymbol{\theta}_{max}$, where the bounds reflect prior knowledge. \cite{byrd1995lbfgs-b} provides the L-BFGS-B algorithmic details for interested readers. 

\subsection{Comparison to existing procedures that estimate $Sr$}

The Diff-GNS framework has fundamental advantages over existing back-analysis techniques explained in \Cref{sec:existing_methods}, addressing their core limitations in physical representation, flexibility, automation, and scalability, as summarized in \Cref{table:model_comparison}.

First, regarding physical fidelity, existing methods rely on simplified assumptions about failure mechanics, such as predefined geometries or slip surfaces. In contrast, the Diff-GNS framework is built upon learned physics of a high-fidelity numerical solver, MPM. This allows Diff-GNS to naturally capture more complex physics-consistent behaviors, while retaining computational efficiency.

Second, Diff-GNS offers greater flexibility in its optimization objective. While all methods (traditional and Diff-GNS) require pre- and post-failure geometries as primary input data, traditional approaches typically target a single, fixed objective, such as matching a final center-of-gravity displacement or achieving a factor of safety of 1.0. The Diff-GNS loss function, however, can be flexibly defined to incorporate a richer set of observational data, including the displacement, full final runout geometry, intermediate runout stages, or even velocity fields.

Lastly, traditional methods rely on an iterative trial-and-error process that require significant engineering judgment, which limits reproducibility. Diff-GNS replaces this subjective workflow with a fully automated, gradient-based optimization pipeline. This automation not only ensures more reproducibility but also makes the framework inherently scalable to high-dimensional inverse problems without an increase in computational cost with respect to the number of parameters.

\begin{sidewaystable}[p]
\centering
\tiny
\renewcommand{\arraystretch}{1.7} 
\resizebox{\textwidth}{!}{%
\begin{tabular}{@{}p{4cm}p{3.5cm}p{3.5cm}p{3.5cm}p{3.5cm}@{}}
\toprule
\textbf{Items} & \textbf{Kinetics method} & \textbf{ZIF} & \textbf{IMM} & \textbf{Diff-GNS} \\
\midrule
\multicolumn{5}{@{}l}{\textbf{A. Physics and geometrical representation}} \\
Governing physics & $F=ma$ on rigid blocks & Pseudo-static analysis at ZIF condition & $F=ma$ at discrete, user-defined incremental geometries & Learned MPM physics \\
Runout Development & Pre-defined polynomial curve & Assumed based on ZIF geometry & User-interpreted & Naturally modeled via learned MPM physics \\
Material interaction & Simplified to rigid blocks & Limited representation at ZIF geometry & Based on assumed geometries & Interaction enabled through learned MPM physics\\
Deformation modeling & Simplified to block sliding & N/A & Based on assumed geometires & Learned MPM physics \\
3D potential & Designed for 2D & Designed for 2D & Designed for 2D & Yes \\
\midrule
\multicolumn{5}{@{}l}{\textbf{B. Data \& input requirements}} \\
Geometry & Pre and post failure geometries & Pre and post failure geometries & Pre, post, and incremental failure geometries & Pre and post failure geometries \\
Objective & Single: match final center of gravity displacement & Single: limit-equilibrium $FS = 1$ at ZIF condition & Multiple: incremental runout stages & Flexible (e.g., displacement, final/intermediate partial or full geometry)\\
\midrule
\multicolumn{5}{@{}l}{\textbf{C. Implementation \& Computation}} \\
Optimization pipeline & Trial and error on $S_r$ & Manual calibration of ZIF geometry and $S_r$ & Manual calibration of incremental geometries and $S_r$ & Differentiable simulator with gradient-based optimization \\
Reproducibility & User dependent & User dependent & User dependent & High: fully automated once GNS is trained\\
Forward evaluation runtime & Efficient & Efficient, but requires iterative ZIF geometry inference& Significant, requires significant time to define incremental geometries & Efficient\\
Multi-parameter scalability & Limited & Limited & Limited & Significant \\
\bottomrule
\end{tabular}%
}
\caption{Features of traditional methods to estimate $S_r$ and comparison to Diff-GNS.}
\label{table:model_comparison}
\end{sidewaystable}

\section{Data and training}\label{sec:training}

Most previous GNS training datasets were designed for proof-of-concept studies with very simplified geometries \citep{sanchez2020learning, choi2024graph, zhao2025physical}, such as a cube-shaped homogeneous material in a 1 × 1 m domains. The only effort targeting realistic slopes, \citet{choi2025differentiable_multilayers}, still relied on geometries not informed by case histories and with limitations in geometric scale representation, i.e., small and large slopes cannot be accommodated effectively. 

\begin{figure}[!htbp]
    \centering
    \includegraphics[width=1.0\textwidth]{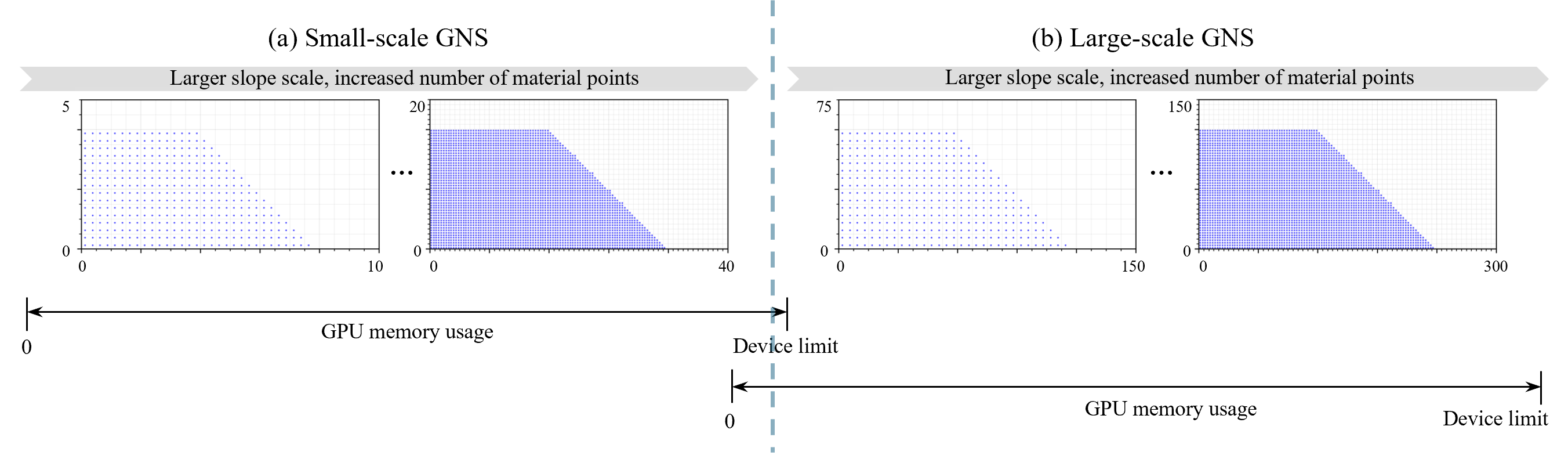}
    \caption{
    Scale-specific separate GNS training strategy. A single GNS cannot efficiently accommodate the wide range of slope dimensions due to resolution and memory constraints: large slopes require an excessive number of material points, exceeding the GPU device's memory limits, while small slopes demand finer material point resolution than a large-scale model can provide. We train separate GNS models for large and small-scale slopes, each with its own representative material point and cell resolution of the underlying MPM training data.
    }
    \label{fig:scale_effect}
\end{figure}

This study addresses these gaps by developing targeted training datasets with geometries informed by runout failure case histories and additional configurations to supplement the training. The datasets expose GNS to various runout mechanisms and adopt a scale-specific strategy (\Cref{fig:scale_effect}) to cover a range of slopes encountered in practice, from a few meters to several hundred meters \citep{olson2001liquefaction_kinetics,weber2015engineering_imm}. A single GNS cannot efficiently handle this spectrum because its resolution is tied to the material point spacing set by the training data’s cell size. Large slopes demand coarser spacing with a larger cell size to stay within GPU memory limits, whereas small slopes require finer spacing for representing adequate detail. We therefore train two complementary GNS models: (i) a large-scale GNS for slopes 200-500 m long and 50-100 m high and (ii) a small-scale GNS for slopes 50-200 m long and 10-50 m high. These ranges are guided by 16 case history failures \citep{olson2001liquefaction_kinetics,weber2015engineering_imm}, summarized in \Cref{table:appendix_dam_dimensions} in \ref{sec:appendix_training_data}. The considered cases serve only to guide the training rather than strict limits of applicability; GNS models can generalize to scenarios outside these ranges, as demonstrated in previous studies \citep{sanchez2020learning,choi2024graph}.

Each model is trained on MPM simulations with cell size and material point resolution tailored to its target scale. \Cref{table:training_datasets} summarizes the MPM configurations for the large- and small-scale datasets. Both designs ensure at least several thousand material points to represent typical case history slopes (\Cref{table:appendix_dam_dimensions}) with sufficient detail, while capping the number of material points at 20,000 to avoid GPU memory overflow (\cite{choi2024inverse}. This dual-scale design mitigates memory bottlenecks, maintains appropriate resolution, and collectively enables proper representation of slope geometries encountered in practice from tens to hundreds of meters---capabilities that prior GNS studies cannot consider.

\begin{figure}[!htbp]
    \centering
    \includegraphics[width=0.8\textwidth]{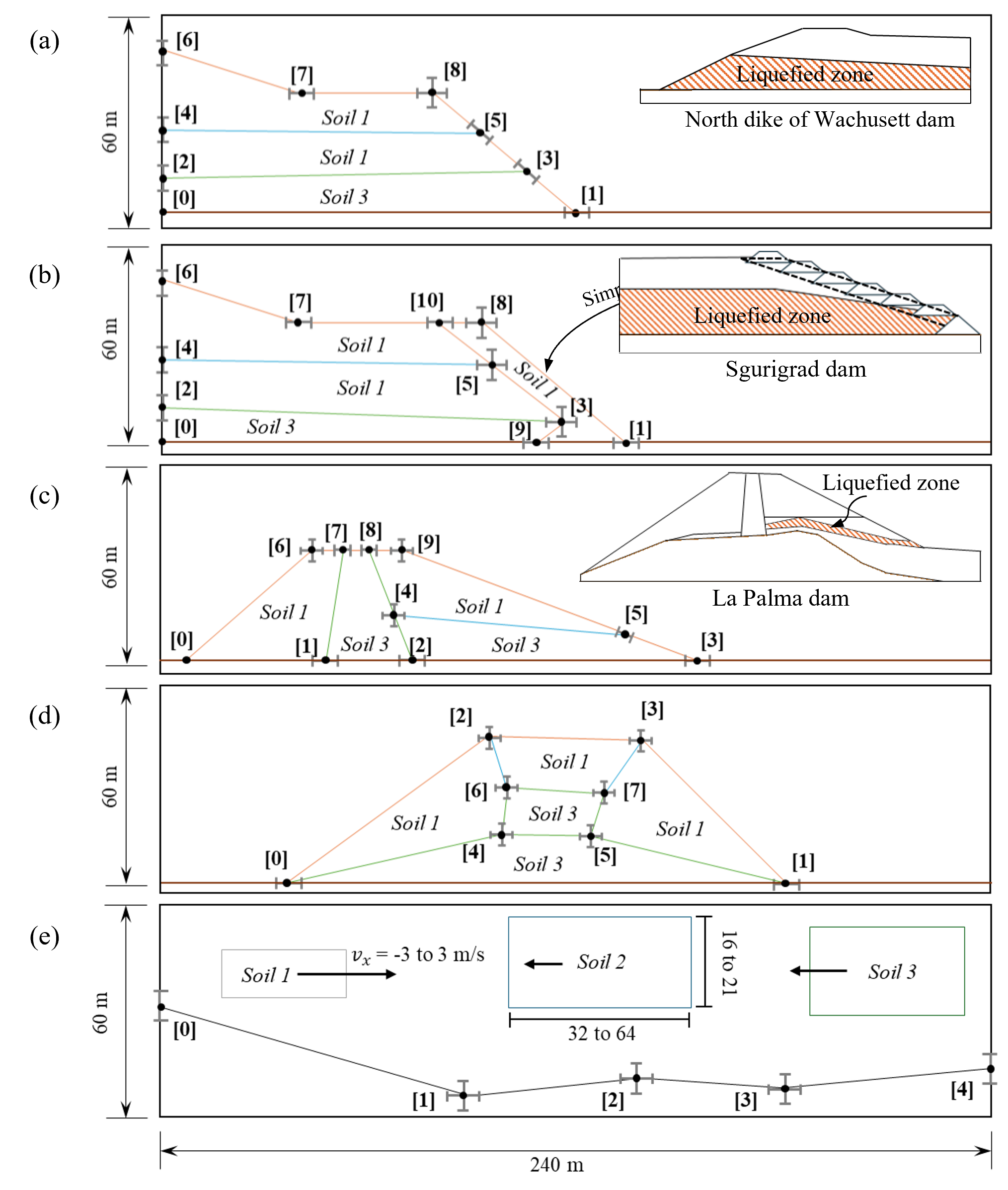}
    \caption{Pre-failure configurations to generate training data for the large-scale GNS. Dimensions for training the small-scale GNS are scaled down by 0.25. The figures are not in scale. (a), (b), and (c) configurations are informed by the case histories listed in \Cref{table:appendix_dam_dimensions} in \ref{sec:appendix_training_data}. Examples are included as figure insets (i.e., North Dike of Wachusett dam, Sgurigrad dam, La Palma dam). (d) and (e) are additional configurations to complement the GNS learning process (see discussions in the text). The numbers in square brackets indicate the indices of the polygon vertices. Solid lines connecting these vertices define the overall shape of granular masses, with brown solid lines denoting the bedrock boundary. The x and y coordinates of the polygon vertices are randomly perturbed; the bars at each vertex represent the perturbation. \Cref{table:appendix_training_data} in \ref{sec:appendix_training_data} provides the variation range for the corresponding vertex indices. Strength properties for Soil 1, Soil 2, and Soil 3 are drawn from distributions in \Cref{table:training_datasets}. In subfigures a-c, the inscribed figures are examples of case histories that we design to emulate. In subfigure e, the black arrow shows the randomly imposed initial velocity.}
    \label{fig:training_data}
\end{figure}

\Cref{fig:training_data} presents the five base pre-failure geometries used to generate the training datasets, with material properties summarized in \Cref{table:training_datasets}. The first three geometries (\Cref{fig:training_data}a-c) are informed by case histories. Note that rather than replicating the geometries, the intent is to distill geometric traits so the GNS can learn representative runout mechanisms. The training is complemented by the last two scenarios to incorporate a variety of mechanisms in the GNS training. The horizontal and vertical bars at each vertex indicate variations applied to the underlying baseline geometry. The variation ranges are tabulated in \Cref{table:appendix_training_data}. These geometric variations introduce diversity into the training dataset.

The first configuration (\Cref{fig:training_data}a) is considered to capture generic rounout features in embankments and some dams (e.g., North Dike of Wachusett dam \citep{olson2000static_wachusett}, Fort Peck dam \citep{marcuson1976dynamic_fort_peck}, Lake Ackerman Highway embankment \citep{hryciw1990liquefaction_lake_ackerman}). As a reference, \Cref{fig:training_data}a shows the geometry of the North Dike of Wachusett dam case history. Regions labeled with the same material index (e.g., soil 1-3) draw their Mohr-Coulomb strength parameters from the same distribution, but do not necessarily share identical properties. These parameters are randomly sampled from uniform distributions in the ranges specified in \Cref{table:training_datasets}. These distributions span the $S_r$ range and non-liquefied strength documented in \citet{olson2001liquefaction_kinetics}. Each geometry contains at least three layers to promote learning of multi-material interactions. Soil 3, representing cohesion-only material, is typically placed in the bottom layers to simulate liquefied zones described by $S_r$.

The second configuration (\Cref{fig:training_data}b) emulates upstream tailings dam geometries (e.g., Sgurigrad dam \citep{international2001tailings}) and their corresponding flow failure. Note that the upstream component is simplified for training purposes as schematized in \Cref{fig:training_data}b. Liquefaction is triggered at the bottom unit to generate runouts.

The third configuration (\Cref{fig:training_data}c) reflects typical zoned dam geometries \citep{clarkson2021tailings_overview}, including distinct core and shell components as observed in case histories (e.g., San Fernando dam \citep{seed1973analysis_lsfd}, La Palma dam \citep{weber2015engineering_imm}),  and simulates runout scenarios after triggering liquefaction on different components. 

The fourth configuration (\Cref{fig:training_data}d) includes random slope configurations subject to runout failure, providing additional diversity to the training set. The final configuration (\Cref{fig:training_data}e) adds a general spectrum of granular flow dynamics, including collision and frictional shearing generated by the rectangular granular bodies subjected to collide with different imposed initial velocities to further inform runout mechanisms to the trained GNS.

The discussed configurations are randomly generated based on \Cref{table:appendix_training_data}. All geometries in \Cref{fig:training_data} are sized for the large-scale GNS; for the small-scale GNS, dimensions are reduced by a factor of four to match the finer material-point spacing required by the smaller MPM cell size.

\begin{table}[]
\centering
\begin{threeparttable}
\caption{Datasets for training large- and small-scale GNSs.}
\label{table:training_datasets}
\begin{tabular}{@{}ccp{4cm}p{4cm}@{}}
\toprule
\multicolumn{2}{c}{\multirow{2}{*}{Property}} & \multicolumn{2}{c}{Dataset} \\ \cmidrule(l){3-4} 
\multicolumn{2}{c}{} & Large-scale & Small-scale \\ \midrule

\multirow{6}{*}{MPM configuration} 
 & Domain boundary & 60 x 240 m & 5 x 30 m \\
 & Cell size & 2 x 2 m & 0.5 x 0.5 m \\
 & Distance between material points & 1 m & 0.25 m \\
 & dt & 5e-5 s & 1e-5 s \\
 & \# material points per cell & \multicolumn{2}{c}{4} \\
 & Max. \# material points & \multicolumn{2}{c}{6,000} \\ 

 \midrule

\multirow{5}{*}{Materials} 
 & Soil 1 & \multicolumn{2}{c}{$\phi = 20 - 47 \degree$} \\
 & Soil 2 & \makecell{$\phi = 20 - 47 \degree$,\\ $c = 20-50 \ kPa$} & \makecell{$\phi = 20 - 47 \degree$,\\ $c = 1-40 \ kPa$} \\
 & Soil 3 & \makecell{$\phi = 0 \degree$,\\ $c = 20-50 \ kPa$} & \makecell{$\phi = 0 \degree$,\\ $c = 1-40 \ kPa$} \\
 & Elastic modulus & \multicolumn{2}{c}{30 MPa} \\
 & Density & \multicolumn{2}{c}{1800 ($kg/m^3$)} \\
 & Poisson ratio & \multicolumn{2}{c}{0.3} \\ \midrule
\multicolumn{2}{c}{\# simulations} & \multicolumn{2}{c}{2,500} \\ \bottomrule
\end{tabular}

\begin{tablenotes}
\footnotesize
\item[a] Uniform distributions are used to sample the $\phi$ and $S_r$ within the ranges tabulated for soils 1 to 3 and assigned to the corresponding regions in \Cref{fig:training_data}.
\end{tablenotes}

\end{threeparttable}
\end{table}

For effective training, we need to properly select the following GNS-specific hyperparameters: connectivity radius, number of message passing steps, and noise standard deviation \citep{sanchez2020learning}. 
The connectivity radius specifies the maximum distance within which two material points are connected by a graph edge for message passing. Larger values capture more local interactions because more neighbors are included in the graph, but increase computational cost. We set the radius to 1.5 times the MPM cell size provided in \Cref{table:training_datasets}. This enables sufficient neighbor connectivity for capturing local interactions in GNS. The number of message passing steps defines how many rounds of message passing occur between connected graph vertices. Increasing this value improves the model’s capacity to capture complex interactions but also raises computational cost. We use 10 steps, which balance the learning capacity and computation cost \citep{choi2024graph,sanchez2020learning}.

In long rollouts, prediction errors accumulate as the model reuses its own outputs as inputs (\Cref{eq:gns}), leading the granular flow state away from the training distribution. To mitigate this error accumulation, we employ the random-walk noise injection strategy of \cite{sanchez2020learning}. During training, Gaussian noise with a specified standard deviation is added cumulatively to material point positions in $\mathbf{x_i^t}$, while the acceleration targets (\Cref{eq:gns_objective}) are adjusted so that the perturbed positions still lead to the correct next state. This improves model stability to error accumulation during rollout. We use a velocity noise standard deviation of 0.067 m/s and 0.020 for large and small GNS training, selected based on the overall velocity standard deviation in the respective training dataset. This noise is then applied cumulatively to perturb material point positions. For further implementation details, refer to \cite{sanchez2020learning} and \cite{pfaff2020learning}. For the other typical deep learning hyperparameters, we use the learning rate of 1e-4 for the Adaptive Moment Estimation (ADAM) optimizer \citep{kingma2014adam}, batch size of 2 with 3 GPUs for the parallel training, yielding the effective batch size of 6. The GPU device is NVIDIA A100, provided by Lonestar 6 at the Texas Advanced Computing Center (TACC).

\begin{figure}[!htbp]
    \centering
    \includegraphics[width=1.0\textwidth]{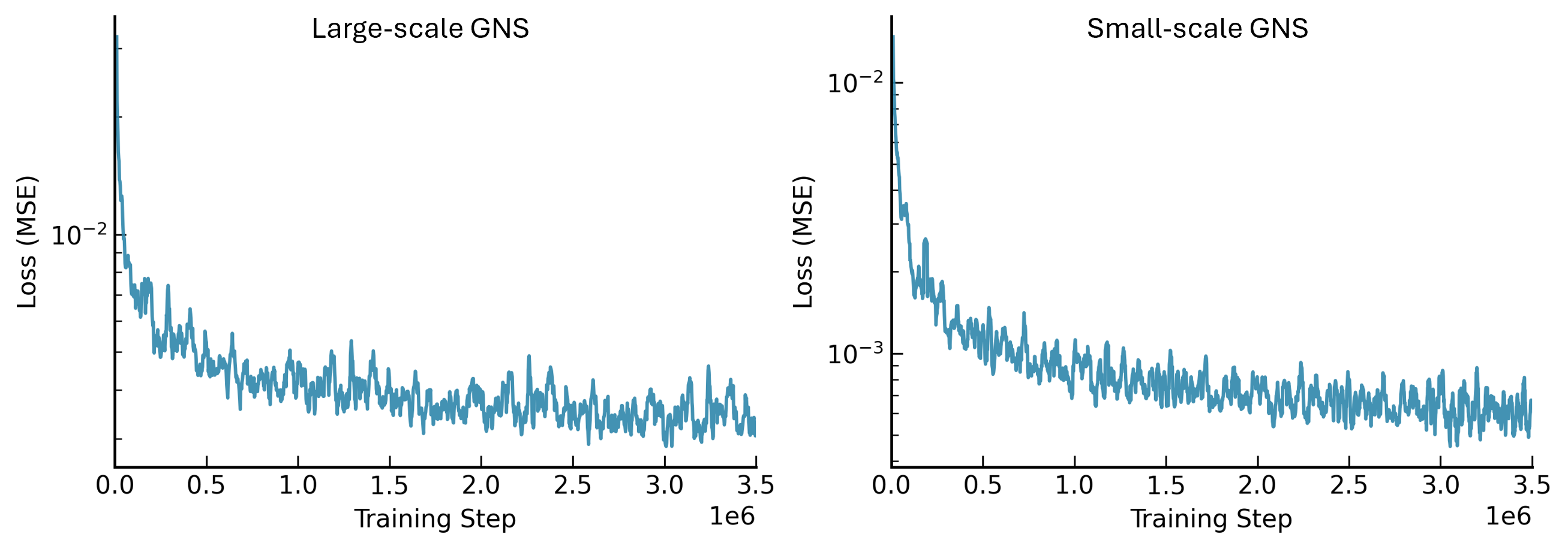}
    \caption{Training loss history.}
    \label{fig:train_history}
\end{figure}

\Cref{fig:train_history} shows the training loss (\Cref{eq:gns_objective}) history for the large and small-scale GNS. In both cases, the mean squared error (MSE) decreases rapidly during the initial training phase, followed by a gradual reduction and stabilization as training progresses. The large-scale GNS converges to a higher final loss compared to the small-scale GNS, since it typically has a higher acceleration level due to the larger system scale. The training time for 3.5 million steps takes about 4 days, corresponding to the end of the GPU runtime on the computing cluster used in this study.

\section{Diff-GNS framework performance in case histories}\label{sec:results}

We evaluate the performance of the Diff-GNS framework for large and small-scale slopes by back-analyzing $S_r$ for two liquefaction-induced flow failure case histories: the Lower San Fernando dam and La Marquesa dam. Brief overviews of these cases are provided in \ref{sec:appendix_case_history}. The discussions in this section focus on the GNS modeling setup and performance assessment. Importantly, none of the case histories, their specific stratigraphy, or material configurations were included in the training dataset; hence, they offer a basis for assessing the Diff-GNS framework on unseen scenarios.

\subsection{Lower San Fernando dam}

\begin{figure}
    \centering
    \includegraphics[width=1.0\linewidth]{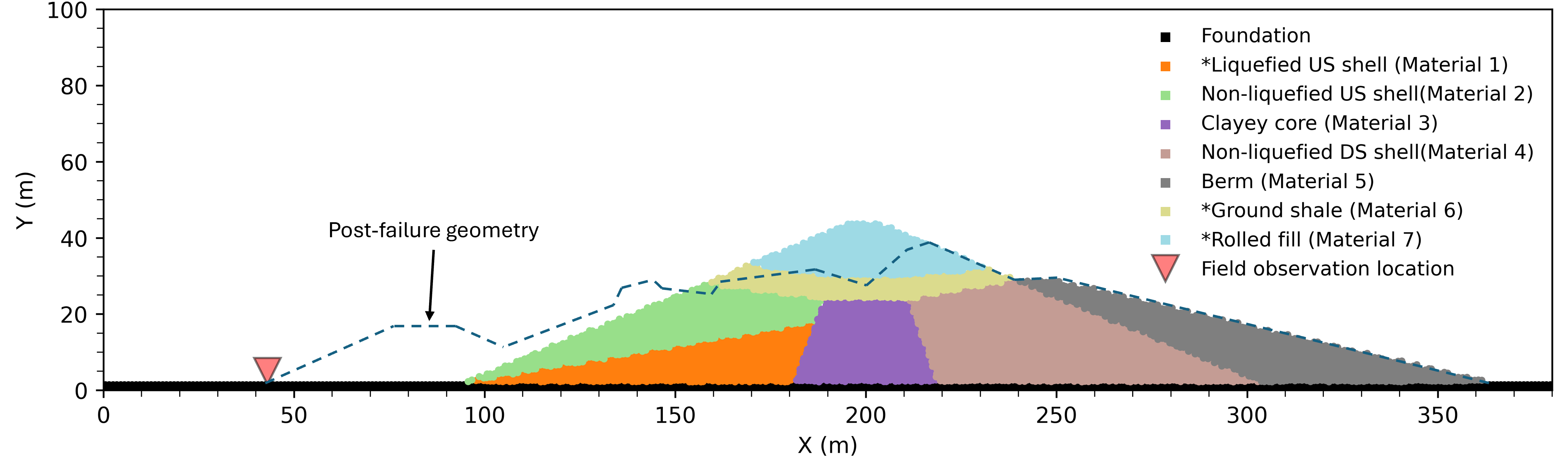}
    \caption{Pre- and post-failure configuration of the Lower San Fernando dam for GNS forward simulation. The asterisk (*) in the legend denotes the materials for which strengths are back-calculated.}
    \label{fig:lsfd_prefailure}
\end{figure}

\Cref{fig:lsfd_prefailure} shows the pre- and post-failure configurations from \cite{weber2015engineering_imm}  for the Lower San Fernando dam used for the GNS simulations. The model contains 6,256 material points. Given the dam’s large size (over 200 m in length), we employ the large-scale GNS model. For simplicity, the drain layer is merged with the berm material (see \Cref{fig:lsfd_section}). Previous studies \citep{olson2001liquefaction_kinetics,weber2015engineering_imm} report uncertainties in the friction angles of ground shale (material 6) and rolled fill (material 7). Hence, the inversion in the Diff-GNS simultaneously estimates the residual strength $S_r$ of material 1 (liquefied upstream hydraulic fill) and the friction angles of materials 6 and 7. The goal is to identify the parameter set that can explain the observed runout. Field observations indicate runouts between 42 m and 65 m \citep{talbot2024modeling}; we adopt 53 m as the target. Initial guesses are $S_r=30$ kPa for material 1 and friction angles of 30$\degree$ for materials 6 and 7, with bounds of 27$\degree$-33$\degree$ and 27$\degree$-37$\degree$, respectively. 
$S_r$ is treated as an operative strength representing the average resistance during the failure process. Unlike earlier back-analyses that only inferred $S_r$, the Diff-GNS can jointly estimate three interacting parameters. Other material properties are summarized in \Cref{table:lsfd_material_properties}.

\begin{figure}[htbp]
    \centering
    
    \begin{subfigure}[b]{0.5\textwidth}  
        \centering
        \includegraphics[width=\textwidth]{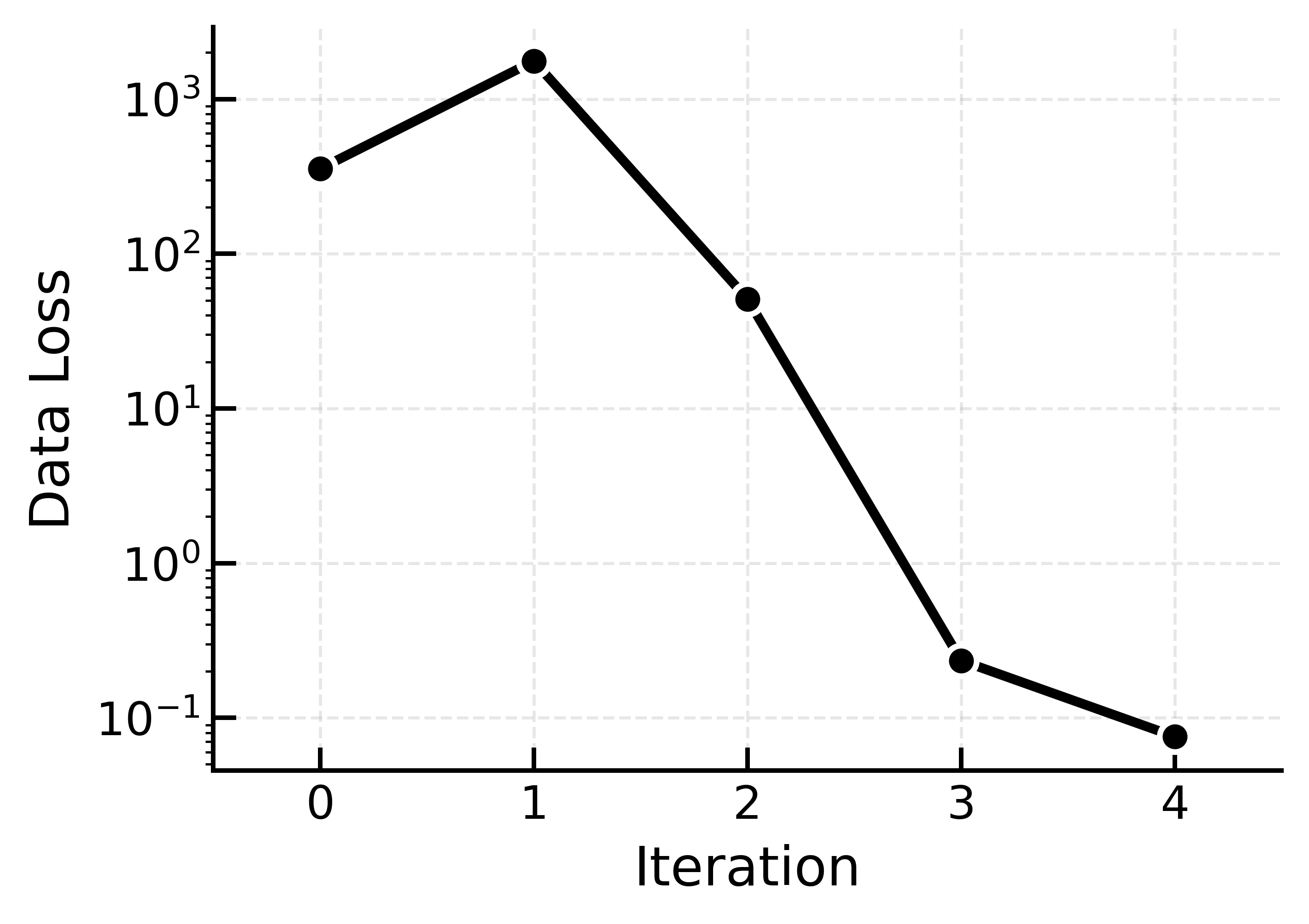}
        \caption{Data loss}
        \label{fig:lsfd_data_loss}
    \end{subfigure}
    
space{1em}  
    
    \begin{subfigure}[b]{1.0\textwidth}  
        \centering
        \includegraphics[width=\textwidth]{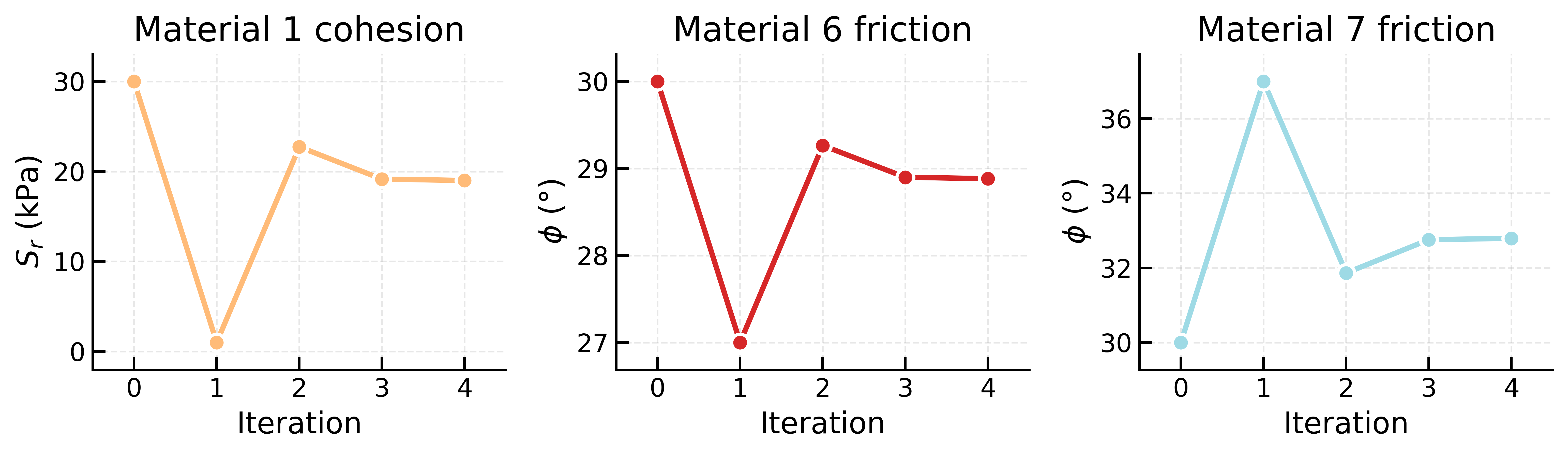}
        \caption{Material parameter history}
        \label{fig:lsfd_param_hist}
    \end{subfigure}

    \caption{Diff-GNS optimization history for the Lower San Fernando dam.}
    \label{fig:lsfd_opt_results}
\end{figure}

\begin{figure}
    \centering
    \includegraphics[width=1.0\linewidth]{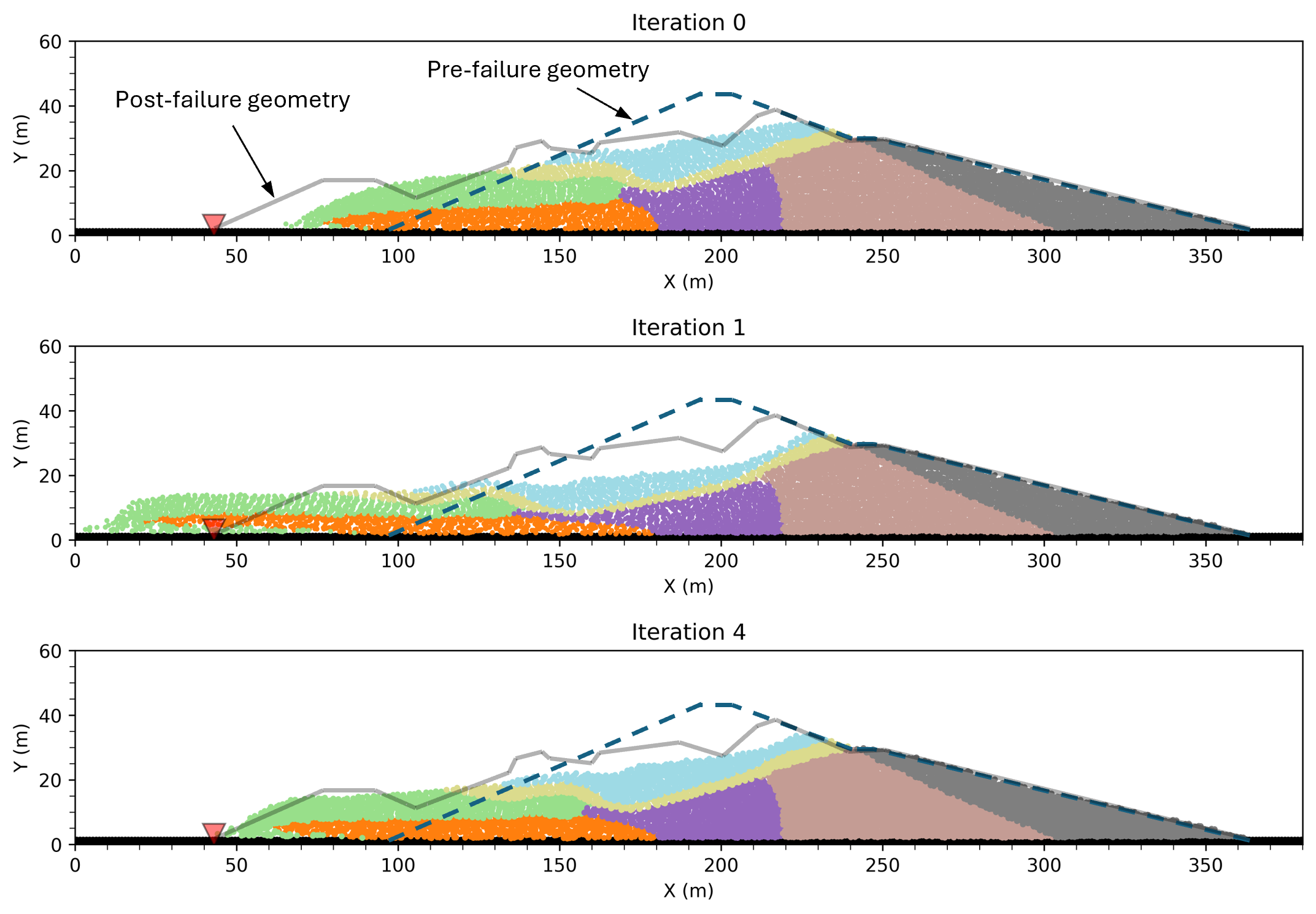}
    \caption{Post-failure geometry evolution throughout the optimization of material parameters for the Lower San Fernando dam. The red inverted triangle indicates the field observed runout. The dashed line represents the pre-failure geometry.}
    \label{fig:lsfd_opt_geom}
\end{figure}

The optimization history (\Cref{fig:lsfd_opt_results}) shows the data loss, defined as the squared error between simulated and observed runout, falling below a pre-defined tolerance of $10^{-1},\mathrm{m}^2$ by the fourth iteration. The inferred $S_r$ for material 1 converges to 18.9 kPa, in close agreement with the 18.7 kPa estimate of \cite{olson2001liquefaction_kinetics}, while the friction angles for materials 6 and 7 converge to 28.9$\degree$ and 32.8$\degree$, respectively (\Cref{fig:lsfd_param_hist}), which are within the uncertainty ranges of 27$\degree$-33$\degree$ and 27$\degree$-37$\degree$for materials 6 and 7 in \citep{olson2001liquefaction_kinetics,weber2015engineering_imm}. The evolving geometries (\Cref{fig:lsfd_opt_geom}) show the simulated runout progressively matching the target runout distance. \Cref{fig:lsfd_opt_geom} illustrates that iteration 0 underestimates the target distance, iteration 1 overestimates it, and by iteration 4, $S_r$ converges to a value that minimizes the data loss. At this stage, the predicted runout geometry is consistent with the target runout and represents the post-failure geometry reasonably well.

The GNS simulation reproduces key deformation features highlighted in \Cref{fig:lsfd_section}. For instance, the upstream shell undergoes major displacement and extrusion, while the downstream shell and berm remain largely intact, and the clay core forms a tongue-like protrusion on the left flank. The crest settles ~20 m, consistent with the value reported by \cite{olson2001liquefaction_kinetics} and \cite{talbot2024modeling}. These patterns mirror prior MPM-based modeling for the Lower San Fernando dam \citep{feng2021mpm-lsfd, talbot2024modeling, tjung2021liquefaction}.

\begin{figure}
    \centering
    \includegraphics[width=1.0\linewidth]{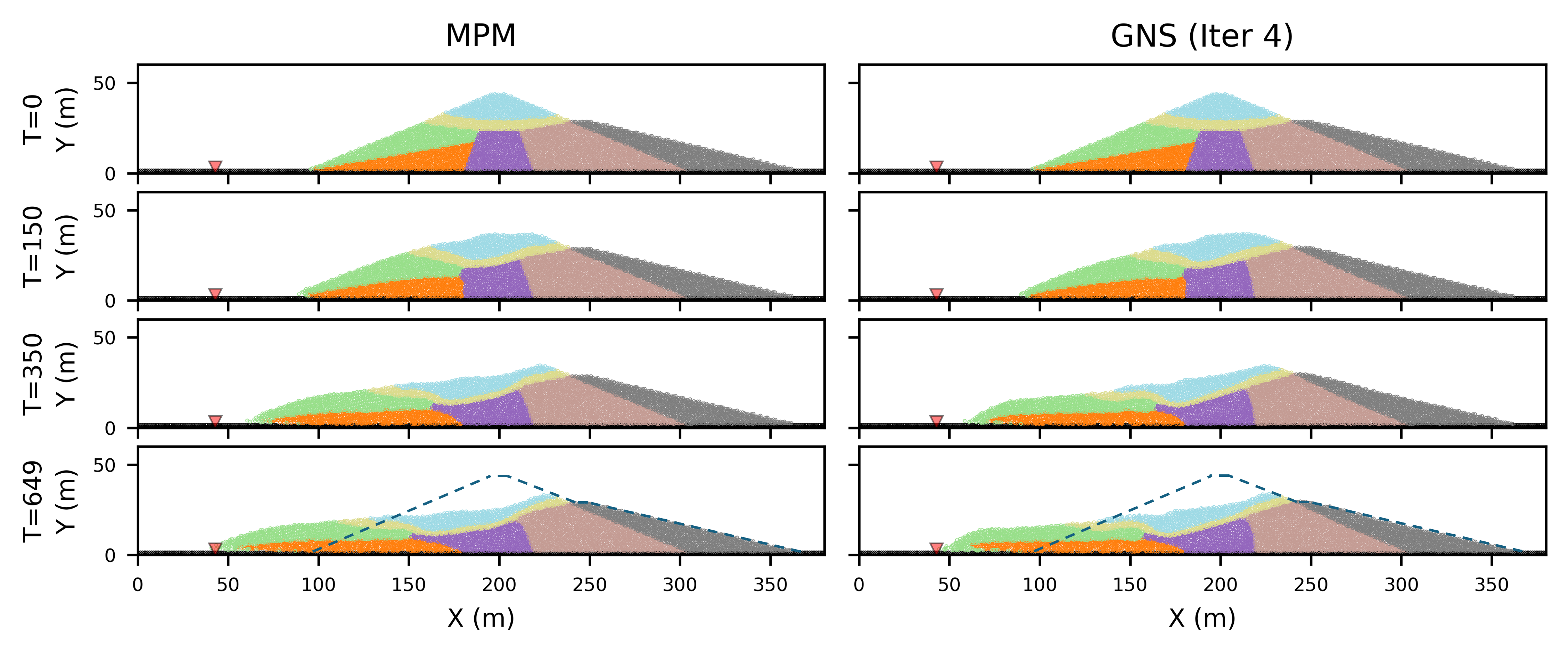}
    \caption{MPM simulation of the Lower San Fernando dam with inferred material properties compared to GNS at optimization iteration 4.}
    \label{fig:lsfd_mpm}
\end{figure}

Finally, to further validate the Diff-GNS results, we ran a high-fidelity MPM simulation using the Diff-GNS-inferred $S_r$ and friction angles. The resulting geometry evolution (\Cref{fig:lsfd_mpm}) closely matches the GNS prediction, including the extent of runout (within $\sim$1 m error) and the characteristic deformation features, confirming that the back-calculated parameters are physically sound.

\begin{table}[h!]
\centering
\resizebox{\textwidth}{!}{%
\begin{tabular}{@{}llcc@{}}
\toprule
\textbf{ID} & \textbf{Description} & $\boldsymbol{\phi}$ (\degree) & $c$ (kPa) \\ 
\midrule
Material 1 & \makecell[l]{Liquefied US shell (back-calculating)} & 0 & 30 (Initial guess) \\
Material 2 & Non-liquefied US shell & 35 & 1 \\
Material 3 & Clayey core & 0 & 30 \\
Material 4 & Non-liquefied DS shell & 35 & 1 \\
Material 5 & Berm & 37 & 0 \\
Material 6 & \makecell[l]{Ground shale (back-calculating with constraint range 27--33\degree)} & 30 (Initial guess) & 1 \\
Material 7 & \makecell[l]{Rolled fill (back-calculating with constraint range 27--33\degree)} & 30 (Initial guess) & 1 \\
\bottomrule
\end{tabular}%
}
\caption{Material properties used in analysis for the Lower San Fernando dam.}
\label{table:lsfd_material_properties}
\end{table}

\subsection{La Marquesa dam}

\begin{figure}
    \centering
    \includegraphics[width=1.0\linewidth]{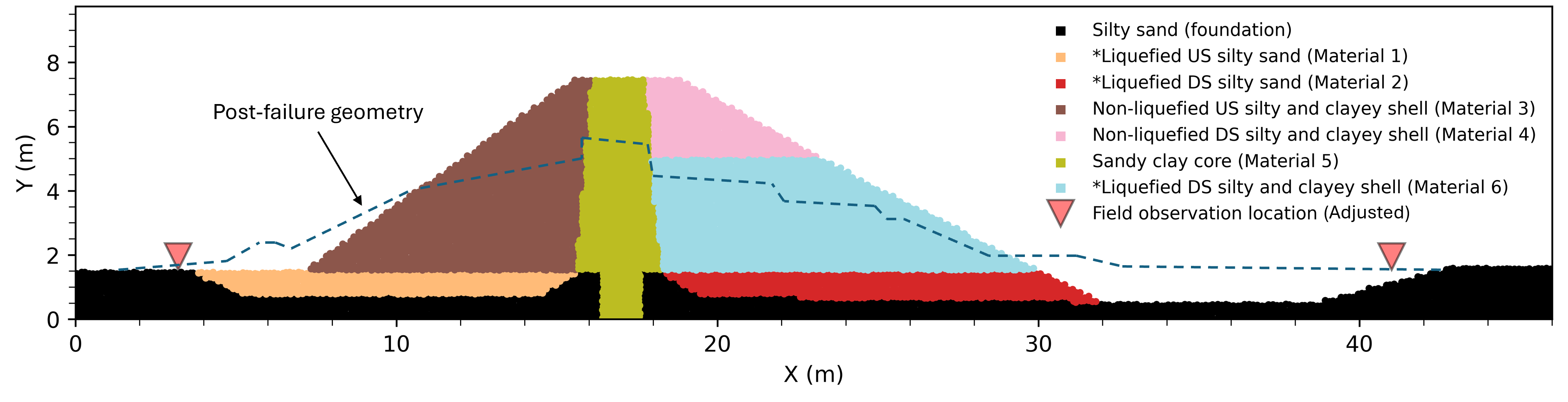}
    \caption{Pre- and post-failure configuration of the La Marquesa dam for GNS forward simulation. The asterisk (*) in the legend denotes the materials for which strengths are back-calculated.}
    \label{fig:lamqsa_prefailure}
\end{figure}

\Cref{fig:lamqsa_prefailure} shows the pre- and post-failure configuration used for the Diff-GNS simulations for the La Marquesa dam, modeled with 8,794 material points. Because the dam is relatively small (length $<$ 50 m), the small-scale GNS model is used. Previous studies \citep{olson2001liquefaction_kinetics,weber2015engineering_imm} treated the liquefied silty sands (materials 1 and 2) and the saturated silty/clayey sand shell (material 6) as a single zone with one residual strength $S_r$. This simplification reflects the difficulty of manually calibrating multiple parameters. In our analyses, Diff-GNS performs a more challenging inversion by independently inferring $S_r$ for materials 1, 2, and 6, which are initialized with $S_r$ of 11 kPa. \Cref{table:lamqsa_material_properties} lists the material properties. The runouts from available post-failure geometries \cite{olson2001liquefaction_kinetics} are estimated as 2.5 m for the upstream slope and 10 m for the downstream slope as indicated by the inverted red triangles in \Cref{fig:lamqsa_prefailure}.

\begin{figure}[htbp]
    \centering
    
    \begin{subfigure}[b]{0.5\textwidth}  
        \centering
        \includegraphics[width=\textwidth]{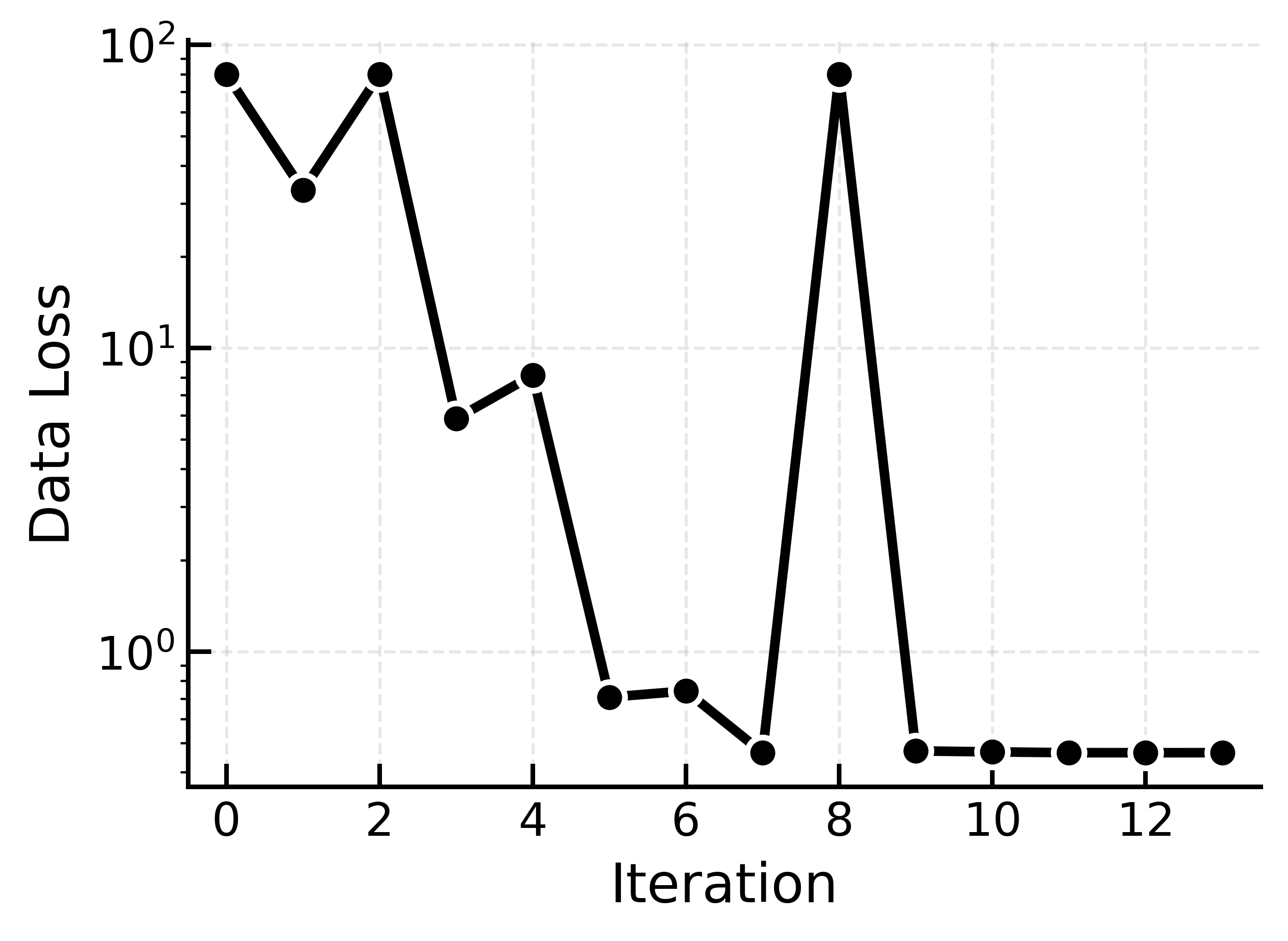}
        \caption{Data loss}
        \label{fig:lamqsa_data_loss}
    \end{subfigure}
    
    
    \begin{subfigure}[b]{1.0\textwidth}  
        \centering
        \includegraphics[width=\textwidth]{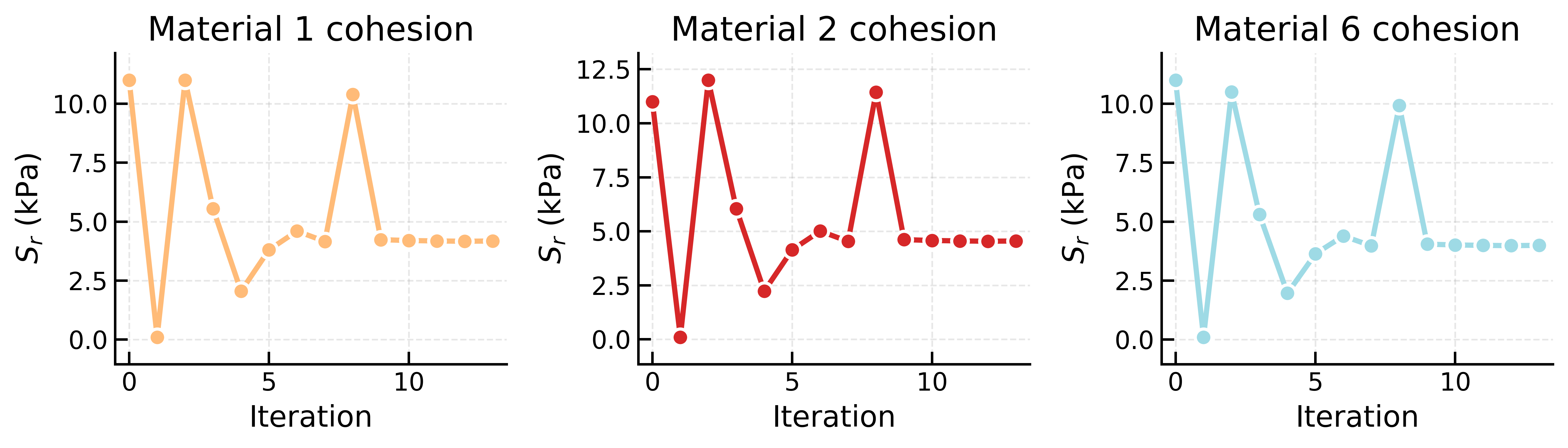}
        \caption{Material parameter history}
        \label{fig:lamqsa_param_hist}
    \end{subfigure}

    \caption{Diff-GNS optimization history for the La Marquesa dam.}
    \label{fig:lamqsa_opt_results}
\end{figure}

\begin{figure}
    \centering
    \includegraphics[width=1.0\linewidth]{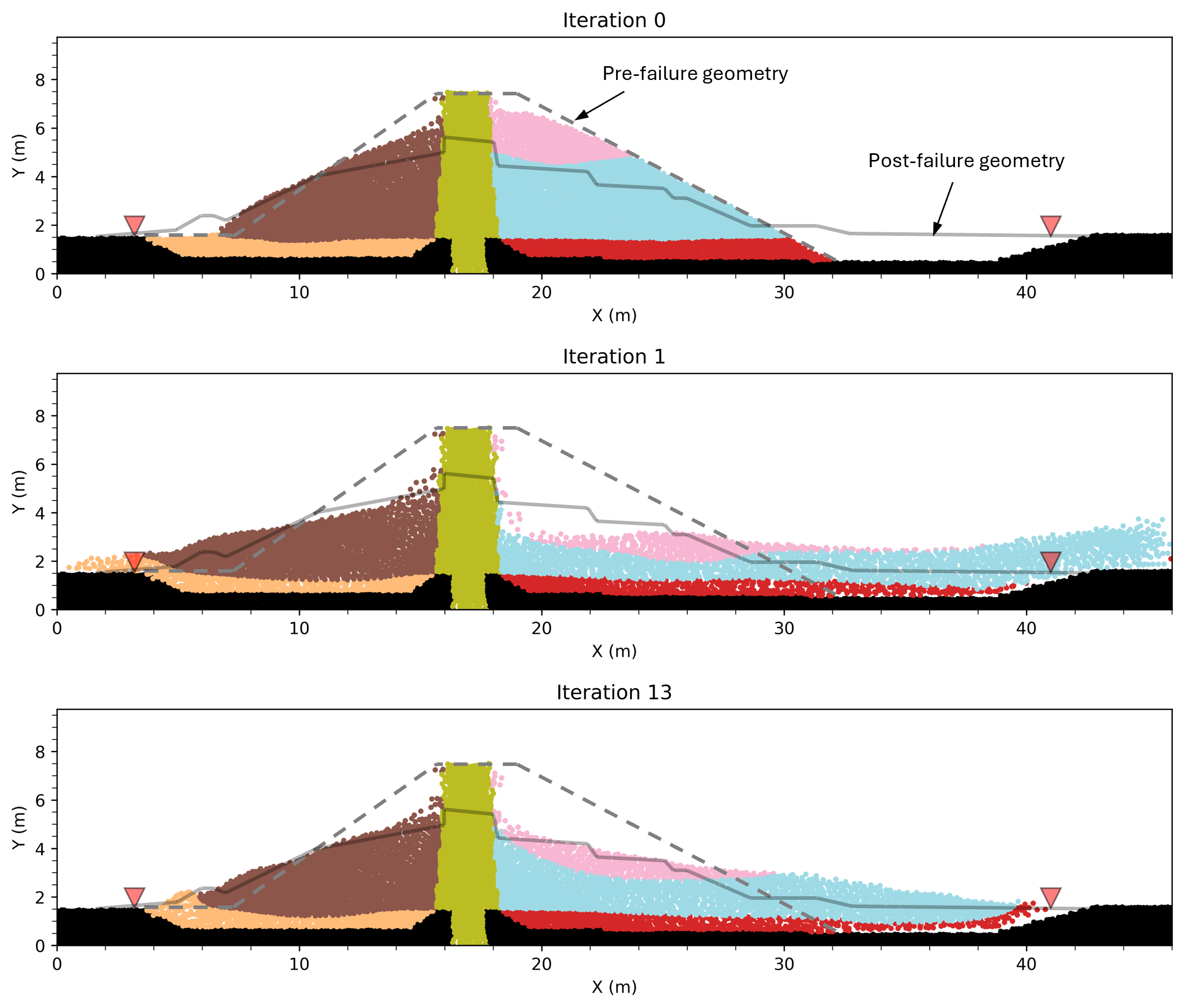}
    \caption{Post-failure geometry evolution throughout the optimization of material parameters for the La Marquesa dam. The red inverted triangle indicates the field observed runout. The dashed line represents the pre-failure geometry.}
    \label{fig:lamqsa_opt_geom}
\end{figure}

\Cref{fig:lamqsa_opt_results} shows the optimization history. The data loss generally decreases and stabilizes by iteration 13, with a temporary increase between iterations 7 and 8. The optimized $S_r$ values converge to 4.2 kPa (material 1), 4.6 kPa (material 2), and 4.00 kPa (material 6), all within the 2.2-9.8 kPa range reported by \cite{olson2001liquefaction_kinetics}, and only slightly below their best estimate of 5 kPa. \Cref{fig:lamqsa_opt_geom} illustrates the runout evolution. At iteration 0, runouts deviate from the targets, and no significant deformations are observed. As optimization proceeds, both upstream and downstream runouts approach the observed targets (red inverted triangles). By iteration 13, the simulated runouts match the observations reasonably well.

The final GNS simulation reproduces the key features of the case history (\Cref{fig:lamqsa_section}), including the extrusion of liquefied materials on both sides and mobilization of the overlying shell. However, it does not replicate the $\sim$2 m settlement of the clay core, likely because its undrained shear strength was assigned as a single value. In field conditions, it is likely that the less confined upper core may have mobilized lower strength and settled more than captured here.

\begin{figure}
    \centering
    \includegraphics[width=1.0\linewidth]{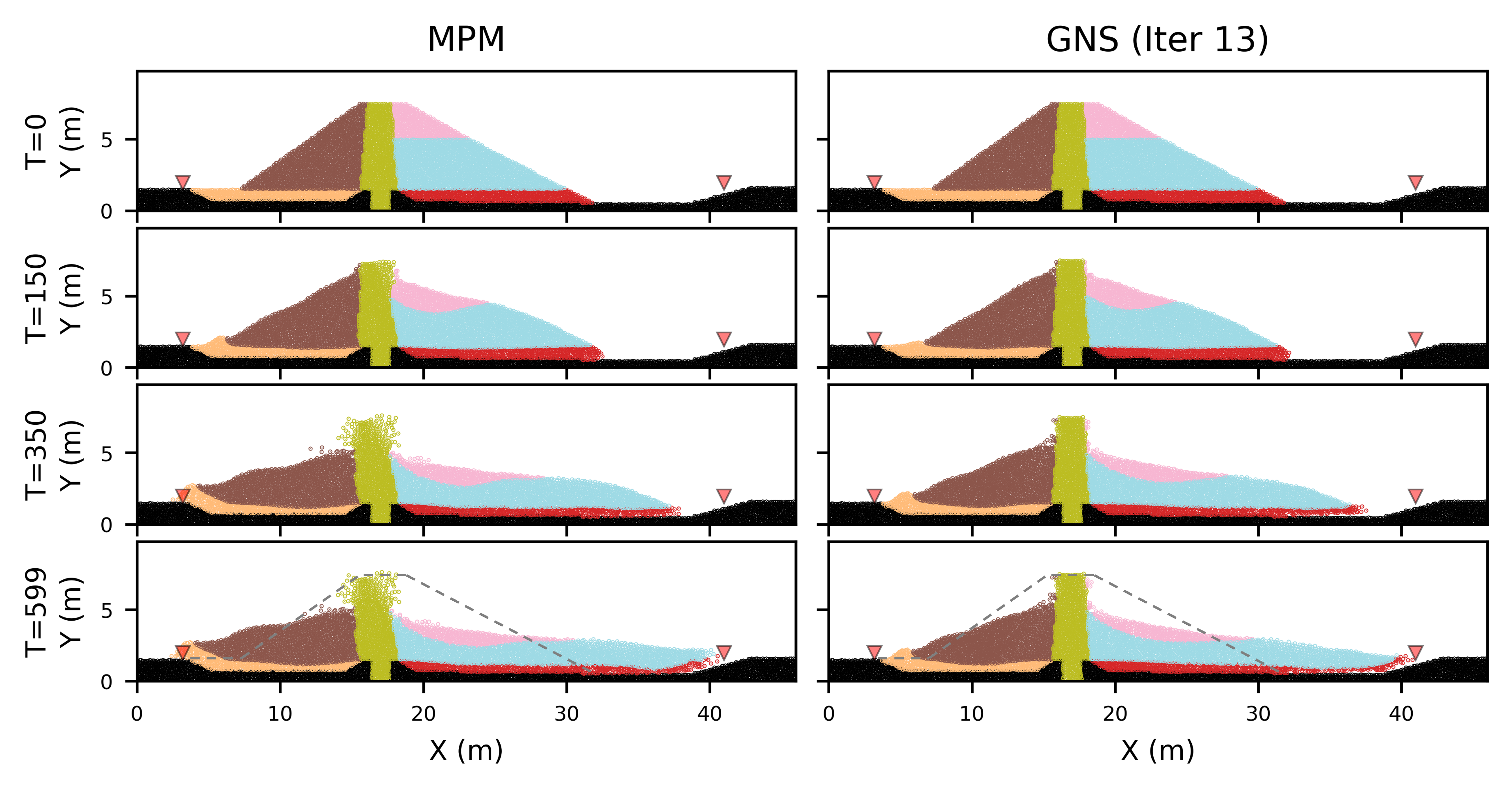}
    \caption{MPM simulation of the La Marquesa dam with inferred material properties compared to GNS at optimization iteration 4.}
    \label{fig:lamqsa_mpm}
\end{figure}

We ran an MPM simulation using the inferred $S_r$ values to further assess the Diff-GNS performance. \Cref{fig:lamqsa_mpm} shows that the MPM results closely follow the Diff-GNS estimations, reproducing the key runout characteristics. Similar to the Diff-GNS, MPM does not capture the vertical core settlement, but it shows a slight leftward core inclination and greater downstream deformation, including more pronounced extrusion of material 1. Despite these minor differences, Diff-GNS efficiently provides physically consistent strength parameters and post-failure runout developments compared to the MPM high-fidelity modeling.

\begin{table}[h!]
\centering
\resizebox{\textwidth}{!}{%
\begin{tabular}{@{}llcc@{}}
\toprule
\textbf{ID} & \textbf{Description} & $\boldsymbol{\phi}$ (\degree) & $c$ (kPa) \\ 
\midrule
Material 1 & \makecell[l]{Liquefied US silty sand (back-calculating)} & 0 & 11 (Initial guess) \\
Material 2 & \makecell[l]{Liquefied DS silty sand (back-calculating)} & 0 & 12 (Initial guess) \\
Material 3 & Non-liquefied DS silty and clayey shell & 30 & 0.1 \\
Material 4 & Non-liquefied US silty and clayey shell & 30 & 0.1 \\
Material 5 & \makecell[l]{Sandy clay core (back-calculating)} & 0 & 24 \\
Material 6 & Liquefied US silty and clayey shell & 0 & 10.5 (Initial guess) \\
\bottomrule
\end{tabular}%
}
\caption{Material properties used in analysis for the La Marquesa dam.}
\label{table:lamqsa_material_properties}
\end{table}

\subsection{Computational efficiency}
\Cref{table:computation_time} shows the computation times for an optimization iteration using Diff-GNS. We also tabulated the MPM forward simulation time for comparison. GNS computation is on RTX 2000 with 16 GB memory, and MPM computation is on Intel Xeon with 56 cores in the TACC Frontera system. The optimization iteration consists of forward simulation and backpropagation to compute the gradient. 

In the case of the Lower San Fernando dam, the MPM forward computation time takes about 1 h, whereas the GNS forward simulation takes about 1 minute, achieving about ×67 speed up. The backpropagation takes about 3 times the forward simulation, i.e., 180 s, so a full optimization loop finishes in about 4 min. For the La Marquesa dam, computation times scale proportionally with the number of material points. The MPM forward simulation takes about 1.2 h, while the GNS forward run requires about 1 min plus 3.4 min for backpropagation, for a total of roughly 4.4 min optimization time.

Diff-GNS computes the multi-dimensional gradient through a single forward simulation with AD. Consequently, the gradient computation time remains constant regardless of the number of parameters, making the framework scalable to high-dimensional inverse problems, i.e., it can be accommodated for inverting a large number of parameters, if needed. In contrast, in MPM, each parameter would require several forward runs, increasing computational costs significantly.

\begin{table}[htbp]
\centering
\caption{Computation time in the Diff-GNS inversion for the considered case histories compared to MPM forward simulation time.}
\resizebox{\textwidth}{!}{%
\begin{tabular}{@{}lrrrrrr@{}}
\toprule
\textbf{Case} & \textbf{\# Material Points} & \textbf{MPM Sim. (s)} & \textbf{\# GNS Steps} & \textbf{GNS Sim. (s)} & \textbf{Gradient (s)} & \textbf{Total (s)} \\
\midrule
Lower San Fernando dam & 6,254 & 3,867 & 700 & 57 & 180 & 237 \\
La Marquesa dam        & 8,794 & 4,331 & 650 & 68 & 204 & 272 \\
\bottomrule
\end{tabular}%
}
\label{table:computation_time}
\end{table}

\section{Conclusions}

This study introduces Diff-GNS as a physics-informed and automated framework for back-calculating post-liquefaction residual strength ($S_r$). The framework addresses key limitations of traditional approaches, which often rely on simplified physics and labor-intensive, subjective workflows. Diff-GNS requires comparable or fewer inputs than traditional approaches \Cref{table:model_comparison} while providing a physically consistent and computationally efficient alternative. It also supports the simultaneous inversion of multiple parameters---capabilities that are difficult to achieve with traditional approaches.

The framework leverages a trained GNS combined with gradient-based optimization through automatic differentiation for the efficient $S_r$ inversion. Addressing limitations from previous GNS-focused efforts, this study trains GNSs for small-scale and large-scale slope systems; the training is informed by case history slope failures and complemented by scenarios that bring a variety of runout mechanisms.

The framework performance for small and large-scale slopes is assessed considering two case histories: the Lower San Fernando and La Marquesa dams. The results show a good performance. Diff-GNS estimated $S_r$ and additional uncertain strength parameters in close agreement with published values. Moreover, it provided a physics-informed runout evolution with a final runout consistent with field observations. Diff-GNS is also significantly more efficient compared to high-fidelity numerical tools such as MPM, which would require several manual iterations to estimate $S_r$. For instance, the GNS forward simulations in this study are approximately 100 times faster than MPM. Based on the evaluated case histories, a complete Diff-GNS inversion iteration (forward simulation and back-propagation) is expected to take between 2 to 4 minutes with a single GPU. The discussed framework and the trained GNSs are also amenable to future refinements. For example, fine-tuning could be conducted in the future with additional training data if required. 

Overall, the work conducted in this study moves Diff-GNS from conceptual validations toward a practical tool for
estimating $Sr$ and simulating runout in slope systems. The discussed framework lays the groundwork for a new paradigm in back-analysis of case histories of slope failures. Future studies further examining the performance of the discussed framework and the trained GNSs for a large number of case histories are encouraged.

\section{Acknowledgment}
This material is based upon work supported by the National Science
Foundation (NSF) under Grant No. 2145092. Any opinions, findings,
conclusions, or recommendations expressed in this material are those of
the author(s) and do not necessarily reflect the views of the National
Science Foundation. The authors acknowledge the Texas Advanced Computing Center (TACC) at The University of Texas at Austin for providing Frontera and Lonestar6 HPC resources to support GNS training (https://www.tacc.utexas.edu).

\section{Competing interests}
The authors declare there are no competing interests.

\section{Availability of data}
Data generated or analyzed during this study are available from the corresponding author upon reasonable request.

\bibliographystyle{elsarticle-harv} 
\bibliography{main}

\begin{thebibliography}{38}
\expandafter\ifx\csname natexlab\endcsname\relax\def\natexlab#1{#1}\fi
\providecommand{\url}[1]{\texttt{#1}}
\providecommand{\href}[2]{#2}
\providecommand{\path}[1]{#1}
\providecommand{\DOIprefix}{doi:}
\providecommand{\ArXivprefix}{arXiv:}
\providecommand{\URLprefix}{URL: }
\providecommand{\Pubmedprefix}{pmid:}
\providecommand{\doi}[1]{\href{http://dx.doi.org/#1}{\path{#1}}}
\providecommand{\Pubmed}[1]{\href{pmid:#1}{\path{#1}}}
\providecommand{\bibinfo}[2]{#2}
\ifx\xfnm\relax \def\xfnm[#1]{\unskip,\space#1}\fi
\bibitem[{Abram et~al.(2022)Abram, Solis, Liang and Kumar}]{abram2022mpm_insitu}
\bibinfo{author}{Abram, G.}, \bibinfo{author}{Solis, A.}, \bibinfo{author}{Liang, Y.}, \bibinfo{author}{Kumar, K.}, \bibinfo{year}{2022}.
\newblock \bibinfo{title}{In situ visualization of regional-scale natural hazards with galaxy and material point method}.
\newblock \bibinfo{journal}{Computing in Science \& Engineering} \bibinfo{volume}{24}, \bibinfo{pages}{31--39}.
\bibitem[{Battaglia et~al.(2018)Battaglia, Hamrick, Bapst, Sanchez-Gonzalez, Zambaldi, Malinowski, Tacchetti, Raposo, Santoro and Faulkner}]{battaglia2018inductive}
\bibinfo{author}{Battaglia, P.W.}, \bibinfo{author}{Hamrick, J.B.}, \bibinfo{author}{Bapst, V.}, \bibinfo{author}{Sanchez-Gonzalez, A.}, \bibinfo{author}{Zambaldi, V.}, \bibinfo{author}{Malinowski, M.}, \bibinfo{author}{Tacchetti, A.}, \bibinfo{author}{Raposo, D.}, \bibinfo{author}{Santoro, A.}, \bibinfo{author}{Faulkner, R.}, \bibinfo{year}{2018}.
\newblock \bibinfo{title}{Relational inductive biases, deep learning, and graph networks}.
\newblock \bibinfo{journal}{arXiv preprint arXiv:1806.01261} .
\bibitem[{Baydin et~al.(2018)Baydin, Pearlmutter, Radul and Siskind}]{baydin2018ad}
\bibinfo{author}{Baydin, A.G.}, \bibinfo{author}{Pearlmutter, B.A.}, \bibinfo{author}{Radul, A.A.}, \bibinfo{author}{Siskind, J.M.}, \bibinfo{year}{2018}.
\newblock \bibinfo{title}{Automatic differentiation in machine learning: a survey}.
\newblock \bibinfo{journal}{Journal of Marchine Learning Research} \bibinfo{volume}{18}, \bibinfo{pages}{1--43}.
\bibitem[{Byrd et~al.(1995)Byrd, Lu, Nocedal and Zhu}]{byrd1995lbfgs-b}
\bibinfo{author}{Byrd, R.H.}, \bibinfo{author}{Lu, P.}, \bibinfo{author}{Nocedal, J.}, \bibinfo{author}{Zhu, C.}, \bibinfo{year}{1995}.
\newblock \bibinfo{title}{A limited memory algorithm for bound constrained optimization}.
\newblock \bibinfo{journal}{SIAM Journal on scientific computing} \bibinfo{volume}{16}, \bibinfo{pages}{1190--1208}.
\bibitem[{Castro et~al.(1992)Castro, Seed, Keller and Seed}]{castro1992steady}
\bibinfo{author}{Castro, G.}, \bibinfo{author}{Seed, R.B.}, \bibinfo{author}{Keller, T.O.}, \bibinfo{author}{Seed, H.B.}, \bibinfo{year}{1992}.
\newblock \bibinfo{title}{Steady-state strength analysis of lower san fernando dam slide}.
\newblock \bibinfo{journal}{Journal of Geotechnical Engineering} \bibinfo{volume}{118}, \bibinfo{pages}{406--427}.
\bibitem[{Ceccato et~al.(2024)Ceccato, Yerro and Di~Carluccio}]{ceccato2024simulating_mpm}
\bibinfo{author}{Ceccato, F.}, \bibinfo{author}{Yerro, A.}, \bibinfo{author}{Di~Carluccio, G.}, \bibinfo{year}{2024}.
\newblock \bibinfo{title}{Simulating landslides with the material point method: Best practices, potentialities, and challenges}.
\newblock \bibinfo{journal}{Engineering Geology} , \bibinfo{pages}{107614}.
\bibitem[{Choi and Kumar(2023)}]{choi2023three}
\bibinfo{author}{Choi, Y.}, \bibinfo{author}{Kumar, K.}, \bibinfo{year}{2023}.
\newblock \bibinfo{title}{Three-dimensional granular flow simulation using graph neural network-based learned simulator}, in: \bibinfo{booktitle}{Geo-Congress 2024}, pp. \bibinfo{pages}{335--344}.
\bibitem[{Choi and Kumar(2024a)}]{choi2024graph}
\bibinfo{author}{Choi, Y.}, \bibinfo{author}{Kumar, K.}, \bibinfo{year}{2024}a.
\newblock \bibinfo{title}{Graph neural network-based surrogate model for granular flows}.
\newblock \bibinfo{journal}{Computers and Geotechnics} \bibinfo{volume}{166}, \bibinfo{pages}{106015}.
\bibitem[{Choi and Kumar(2024b)}]{choi2024inverse}
\bibinfo{author}{Choi, Y.}, \bibinfo{author}{Kumar, K.}, \bibinfo{year}{2024}b.
\newblock \bibinfo{title}{Inverse analysis of granular flows using differentiable graph neural network simulator}.
\newblock \bibinfo{journal}{Computers and Geotechnics} \bibinfo{volume}{171}, \bibinfo{pages}{106374}.
\bibitem[{Choi et~al.(2025)Choi, Macedo and Liu}]{choi2025differentiable_multilayers}
\bibinfo{author}{Choi, Y.}, \bibinfo{author}{Macedo, J.}, \bibinfo{author}{Liu, C.}, \bibinfo{year}{2025}.
\newblock \bibinfo{title}{Differentiable graph neural network simulator for forward and inverse modeling of multi-layered slope system with multiple material properties}.
\newblock \bibinfo{journal}{arXiv preprint arXiv:2504.15938} .
\bibitem[{Chowdhury(2018)}]{chowdhury2018evaluation}
\bibinfo{author}{Chowdhury, K.H.}, \bibinfo{year}{2018}.
\newblock \bibinfo{title}{Evaluation of the state of practice regarding nonlinear seismic deformation analyses of embankment dams subject to soil liquefaction based on case histories}.
\newblock \bibinfo{publisher}{University of California, Berkeley}.
\bibitem[{Clarkson and Williams(2021)}]{clarkson2021tailings_overview}
\bibinfo{author}{Clarkson, L.}, \bibinfo{author}{Williams, D.}, \bibinfo{year}{2021}.
\newblock \bibinfo{title}{An overview of conventional tailings dam geotechnical failure mechanisms}.
\newblock \bibinfo{journal}{Mining, Metallurgy \& Exploration} \bibinfo{volume}{38}, \bibinfo{pages}{1305--1328}.
\bibitem[{De~Alba et~al.(1988)De~Alba, Seed, Retamal and Seed}]{de1988analyses_lamqsa}
\bibinfo{author}{De~Alba, P.A.}, \bibinfo{author}{Seed, H.B.}, \bibinfo{author}{Retamal, E.}, \bibinfo{author}{Seed, R.B.}, \bibinfo{year}{1988}.
\newblock \bibinfo{title}{Analyses of dam failures in 1985 chilean earthquake}.
\newblock \bibinfo{journal}{Journal of Geotechnical Engineering} \bibinfo{volume}{114}, \bibinfo{pages}{1414--1434}.
\bibitem[{De~Vaucorbeil et~al.(2020)De~Vaucorbeil, Nguyen, Sinaie and Wu}]{de2020mpm}
\bibinfo{author}{De~Vaucorbeil, A.}, \bibinfo{author}{Nguyen, V.P.}, \bibinfo{author}{Sinaie, S.}, \bibinfo{author}{Wu, J.Y.}, \bibinfo{year}{2020}.
\newblock \bibinfo{title}{Material point method after 25 years: Theory, implementation, and applications}.
\newblock \bibinfo{journal}{Advances in applied mechanics} \bibinfo{volume}{53}, \bibinfo{pages}{185--398}.
\bibitem[{Feng et~al.(2021)Feng, Wang, Huang and Jin}]{feng2021mpm-lsfd}
\bibinfo{author}{Feng, K.}, \bibinfo{author}{Wang, G.}, \bibinfo{author}{Huang, D.}, \bibinfo{author}{Jin, F.}, \bibinfo{year}{2021}.
\newblock \bibinfo{title}{Material point method for large-deformation modeling of coseismic landslide and liquefaction-induced dam failure}.
\newblock \bibinfo{journal}{Soil Dynamics and Earthquake Engineering} \bibinfo{volume}{150}, \bibinfo{pages}{106907}.
\bibitem[{Hryciw et~al.(1990)Hryciw, Vitton and Thomann}]{hryciw1990liquefaction_lake_ackerman}
\bibinfo{author}{Hryciw, R.D.}, \bibinfo{author}{Vitton, S.}, \bibinfo{author}{Thomann, T.G.}, \bibinfo{year}{1990}.
\newblock \bibinfo{title}{Liquefaction and flow failure during seismic exploration}.
\newblock \bibinfo{journal}{Journal of Geotechnical Engineering} \bibinfo{volume}{116}, \bibinfo{pages}{1881--1899}.
\bibitem[{Kingma(2014)}]{kingma2014adam}
\bibinfo{author}{Kingma, D.P.}, \bibinfo{year}{2014}.
\newblock \bibinfo{title}{Adam: A method for stochastic optimization}.
\newblock \bibinfo{journal}{arXiv preprint arXiv:1412.6980} .
\bibitem[{Kramer and Wang(2015)}]{kramer2015empirical_zif}
\bibinfo{author}{Kramer, S.L.}, \bibinfo{author}{Wang, C.H.}, \bibinfo{year}{2015}.
\newblock \bibinfo{title}{Empirical model for estimation of the residual strength of liquefied soil}.
\newblock \bibinfo{journal}{Journal of Geotechnical and Geoenvironmental Engineering} \bibinfo{volume}{141}, \bibinfo{pages}{04015038}.
\bibitem[{on~Large~Dams(2001)}]{international2001tailings}
\bibinfo{author}{on~Large~Dams, I.C.}, \bibinfo{year}{2001}.
\newblock \bibinfo{title}{Tailings dams: risk of dangerous occurrences: lessons learnt from practical experiences}.
\newblock \bibinfo{number}{121}, \bibinfo{publisher}{United Nations Publications}.
\bibitem[{Macedo et~al.(2024a)Macedo, Yerro, Cornejo and Pierce}]{macedo2024mpm-granular}
\bibinfo{author}{Macedo, J.}, \bibinfo{author}{Yerro, A.}, \bibinfo{author}{Cornejo, R.}, \bibinfo{author}{Pierce, I.}, \bibinfo{year}{2024}a.
\newblock \bibinfo{title}{Cadia tsf failure assessment considering triggering and posttriggering mechanisms}.
\newblock \bibinfo{journal}{Journal of Geotechnical and Geoenvironmental Engineering} \bibinfo{volume}{150}, \bibinfo{pages}{04024011}.
\bibitem[{Macedo et~al.(2024b)Macedo, Yerro, Cornejo and Pierce}]{macedo2024runout_mpm_cadia}
\bibinfo{author}{Macedo, J.}, \bibinfo{author}{Yerro, A.}, \bibinfo{author}{Cornejo, R.}, \bibinfo{author}{Pierce, I.}, \bibinfo{year}{2024}b.
\newblock \bibinfo{title}{Cadia tsf failure assessment considering triggering and posttriggering mechanisms}.
\newblock \bibinfo{journal}{Journal of Geotechnical and Geoenvironmental Engineering} \bibinfo{volume}{150}, \bibinfo{pages}{04024011}.
\bibitem[{Marcuson and Krinitzsky(1976)}]{marcuson1976dynamic_fort_peck}
\bibinfo{author}{Marcuson, W.F.}, \bibinfo{author}{Krinitzsky, E.}, \bibinfo{year}{1976}.
\newblock \bibinfo{title}{Dynamic analysis of Fort Peck dam}.
\newblock \bibinfo{publisher}{US Department of Defense, Department of the Army, Corps of Engineers~…}.
\bibitem[{Moss et~al.(2019)Moss, Gebhart, Frost and Ledezma}]{moss2019flow}
\bibinfo{author}{Moss, R.}, \bibinfo{author}{Gebhart, T.}, \bibinfo{author}{Frost, D.}, \bibinfo{author}{Ledezma, C.}, \bibinfo{year}{2019}.
\newblock \bibinfo{title}{Flow-failure case history of the las palmas, chile, tailings dam}.
\newblock \bibinfo{journal}{Pacific Earthquake Engineering Research Center PEER report} \bibinfo{volume}{1}.
\bibitem[{Olson(2001)}]{olson2001liquefaction_kinetics}
\bibinfo{author}{Olson, S.M.}, \bibinfo{year}{2001}.
\newblock \bibinfo{title}{Liquefaction analysis of level and sloping ground using field case histories and penetration resistance}.
\newblock \bibinfo{publisher}{University of Illinois at Urbana-Champaign}.
\bibitem[{Olson and Stark(2002)}]{olson2002liquefied}
\bibinfo{author}{Olson, S.M.}, \bibinfo{author}{Stark, T.D.}, \bibinfo{year}{2002}.
\newblock \bibinfo{title}{Liquefied strength ratio from liquefaction flow failure case histories}.
\newblock \bibinfo{journal}{Canadian Geotechnical Journal} \bibinfo{volume}{39}, \bibinfo{pages}{629--647}.
\bibitem[{Olson et~al.(2000)Olson, Stark, Walton and Castro}]{olson2000static_wachusett}
\bibinfo{author}{Olson, S.M.}, \bibinfo{author}{Stark, T.D.}, \bibinfo{author}{Walton, W.H.}, \bibinfo{author}{Castro, G.}, \bibinfo{year}{2000}.
\newblock \bibinfo{title}{1907 static liquefaction flow failure of the north dike of wachusett dam}.
\newblock \bibinfo{journal}{Journal of Geotechnical and Geoenvironmental Engineering} \bibinfo{volume}{126}, \bibinfo{pages}{1184--1193}.
\newblock \DOIprefix\doi{10.1061/(ASCE)1090-0241(2000)126:12(1184)}.
\bibitem[{Pfaff et~al.(2020)Pfaff, Fortunato, Sanchez-Gonzalez and Battaglia}]{pfaff2020learning}
\bibinfo{author}{Pfaff, T.}, \bibinfo{author}{Fortunato, M.}, \bibinfo{author}{Sanchez-Gonzalez, A.}, \bibinfo{author}{Battaglia, P.}, \bibinfo{year}{2020}.
\newblock \bibinfo{title}{Learning mesh-based simulation with graph networks}, in: \bibinfo{booktitle}{International conference on learning representations}.
\bibitem[{Poulos et~al.(1985)Poulos, Castro and France}]{poulos1985liquefaction}
\bibinfo{author}{Poulos, S.J.}, \bibinfo{author}{Castro, G.}, \bibinfo{author}{France, J.W.}, \bibinfo{year}{1985}.
\newblock \bibinfo{title}{Liquefaction evaluation procedure}.
\newblock \bibinfo{journal}{Journal of Geotechnical Engineering} \bibinfo{volume}{111}, \bibinfo{pages}{772--792}.
\bibitem[{Sanchez-Gonzalez et~al.(2020)Sanchez-Gonzalez, Godwin, Pfaff, Ying, Leskovec and Battaglia}]{sanchez2020learning}
\bibinfo{author}{Sanchez-Gonzalez, A.}, \bibinfo{author}{Godwin, J.}, \bibinfo{author}{Pfaff, T.}, \bibinfo{author}{Ying, R.}, \bibinfo{author}{Leskovec, J.}, \bibinfo{author}{Battaglia, P.}, \bibinfo{year}{2020}.
\newblock \bibinfo{title}{Learning to simulate complex physics with graph networks}, in: \bibinfo{booktitle}{International conference on machine learning}, \bibinfo{organization}{PMLR}. pp. \bibinfo{pages}{8459--8468}.
\bibitem[{Seed et~al.(1973)Seed, Lee, Idriss and Makdisi}]{seed1973analysis_lsfd}
\bibinfo{author}{Seed, H.}, \bibinfo{author}{Lee, K.}, \bibinfo{author}{Idriss, I.}, \bibinfo{author}{Makdisi, F.}, \bibinfo{year}{1973}.
\newblock \bibinfo{title}{Analysis of the slides in the san fernando dams during the earthquake of february 9, 1971, report no. eerc 73-2}.
\newblock \bibinfo{journal}{Earthquake Engineering Research Center, Univ. of California, Berkeley, Calif} .
\bibitem[{Seed(1987)}]{seed1987design}
\bibinfo{author}{Seed, H.B.}, \bibinfo{year}{1987}.
\newblock \bibinfo{title}{Design problems in soil liquefaction}.
\newblock \bibinfo{journal}{Journal of Geotechnical Engineering} \bibinfo{volume}{113}, \bibinfo{pages}{827--845}.
\bibitem[{Seed et~al.(1975)Seed, Idriss, Lee and Makdisi}]{seed1975dynamic_analysis_lsfd}
\bibinfo{author}{Seed, H.B.}, \bibinfo{author}{Idriss, I.M.}, \bibinfo{author}{Lee, K.L.}, \bibinfo{author}{Makdisi, F.I.}, \bibinfo{year}{1975}.
\newblock \bibinfo{title}{Dynamic analysis of the slide in the lower san fernando dam during the earthquake of february 9, 1971}.
\newblock \bibinfo{journal}{Journal of the Geotechnical Engineering division} \bibinfo{volume}{101}, \bibinfo{pages}{889--911}.
\bibitem[{Seed et~al.(1989)Seed, Seed, Harder and Jong}]{seed1989re_lsfd}
\bibinfo{author}{Seed, H.B.}, \bibinfo{author}{Seed, R.B.}, \bibinfo{author}{Harder, L.F.}, \bibinfo{author}{Jong, H.L.}, \bibinfo{year}{1989}.
\newblock \bibinfo{title}{Re-Evaluation of the Lower San Fernando Dam; Report 2: Examination of the Post-Earthquake Slide of February 9, 1971}.
\newblock \bibinfo{type}{Technical Report}. Waterways Experiment Station (U.S.).
\bibitem[{Seed and Harder~Jr(1990)}]{seed1990spt}
\bibinfo{author}{Seed, R.}, \bibinfo{author}{Harder~Jr, L.}, \bibinfo{year}{1990}.
\newblock \bibinfo{title}{Spt-based analysis of cyclic pore pressure generation and undrained residual strength”: Proc., hb seed memorial symp., vol. 2}.
\bibitem[{Talbot et~al.(2024)Talbot, Given, Tjung, Liang, Chowdhury, Seed and Soga}]{talbot2024modeling}
\bibinfo{author}{Talbot, L.E.}, \bibinfo{author}{Given, J.}, \bibinfo{author}{Tjung, E.Y.}, \bibinfo{author}{Liang, Y.}, \bibinfo{author}{Chowdhury, K.}, \bibinfo{author}{Seed, R.}, \bibinfo{author}{Soga, K.}, \bibinfo{year}{2024}.
\newblock \bibinfo{title}{Modeling large-deformation features of the lower san fernando dam failure with the material point method}.
\newblock \bibinfo{journal}{Computers and Geotechnics} \bibinfo{volume}{165}, \bibinfo{pages}{105881}.
\bibitem[{Tjung and Soga(2021)}]{tjung2021liquefaction}
\bibinfo{author}{Tjung, E.}, \bibinfo{author}{Soga, K.}, \bibinfo{year}{2021}.
\newblock \bibinfo{title}{Liquefaction-induced dam failure simulation--a case for the material point method}.
\newblock \bibinfo{journal}{arXiv preprint arXiv:2111.13584} .
\bibitem[{Weber(2015)}]{weber2015engineering_imm}
\bibinfo{author}{Weber, J.P.}, \bibinfo{year}{2015}.
\newblock \bibinfo{title}{Engineering evaluation of post-liquefaction strength}.
\newblock \bibinfo{publisher}{University of California, Berkeley}.
\bibitem[{Zhao et~al.(2025)Zhao, Chen and Zhao}]{zhao2025physical}
\bibinfo{author}{Zhao, S.}, \bibinfo{author}{Chen, H.}, \bibinfo{author}{Zhao, J.}, \bibinfo{year}{2025}.
\newblock \bibinfo{title}{A physical-information-flow-constrained temporal graph neural network-based simulator for granular materials}.
\newblock \bibinfo{journal}{Computer Methods in Applied Mechanics and Engineering} \bibinfo{volume}{433}, \bibinfo{pages}{117536}.

\end{thebibliography}

\clearpage
\appendix

\setcounter{figure}{0}
\setcounter{table}{0}
\renewcommand{\thefigure}{A.\arabic{figure}}
\renewcommand{\thetable}{A.\arabic{table}}

\section{Training data}\label{sec:appendix_training_data}

\begin{table}[!htbp]
\footnotesize
\centering
\caption{Coordinates of the polygon points in \Cref{fig:training_data}.}
\setlength{\tabcolsep}{10pt}
\renewcommand{\arraystretch}{0.95}
\begin{tabular}{@{}lll@{}}
\toprule
Point ID & X (m) & Y (m) \\ \midrule

\multicolumn{3}{@{}l}{\textbf{\Cref{fig:training_data}a}} \\
{[0]} & 0 & 2 \\
{[1]} & 120 to 130 & 2 \\
{[2]} & 0 & 17 \\
{[3]} & 1/3 interpolation between {[1]} and {[8]} & 1/3 interpolation between {[1]} and {[8]} \\
{[4]} & 0 & 34 \\
{[5]} & 2/3 interpolation between {[1]} and {[8]} & 2/3 interpolation between {[1]} and {[8]} \\
{[6]} & 0 & Y of {[8]} + (–2.5 to 10) \\
{[7]} & 70 to 80 & Y of {[8]} \\
{[8]} & 70 to 80 & 37 to 47 \\

\multicolumn{3}{@{}l}{\textbf{\Cref{fig:training_data}b}} \\
{[0]} & 0 & 2 \\
{[1]} & X of {[1]} – (20 to 30) & 2 \\
{[2]} & 0 & 17 \\
{[3]} & 1/3 interpolation between {[1]} and {[8]} – (5 to 10) & 1/3 interpolation between {[1]} and {[8]} \\
{[4]} & 0 & 32 \\
{[5]} & 2/3 interpolation between {[1]} and {[8]} – (5 to 10) & 2/3 interpolation between {[1]} and {[8]} \\
{[6]} & 0 & Y of {[8]} + (–5 to 20) \\
{[7]} & 35 to 40 & Y of {[8]} \\
{[8]} & 70 to 80 & 37 to 47 \\
{[9]} & X of [1] – (10 to 15) & 2 \\
{[10]} & X of [8] – (5 to 10) & Y of [8] \\

\multicolumn{3}{@{}l}{\textbf{\Cref{fig:training_data}c}} \\
{[0]} & 5 & 2 \\
{[1]} & 0.2 to 0.3 interpolation between {[0]} and {[3]} & 2 \\
{[2]} & 0.5 to 0.6 interpolation between {[0]} and {[3]} & 2 \\
{[3]} & 135 to 180 & 2 \\
{[4]} & 0.4 to 0.6 interpolation between {[8]} and {[2]} & 0.4 to 0.6 interpolation between {[8]} and {[2]} \\
{[5]} & 0 to 0.6 interpolation between {[3]} and {[9]} & 0 to 0.6 interpolation between {[3]} and {[9]} \\
{[6]} & X of {[0]} + (40 to 50) & 35 to 50 \\
{[7]} & 0.1 to 0.45 interpolation between {[6]} and {[9]} & 0.1 to 0.45 interpolation between {[6]} and {[9]} \\
{[8]} & 0.5 to 0.9 interpolation between {[6]} and {[9]} & 0.5 to 0.9 interpolation between {[6]} and {[9]} \\
{[9]} & 75 to 90 & Y of {[6]} \\

\multicolumn{3}{@{}l}{\textbf{\Cref{fig:training_data}d}} \\
{[0]} & 50 to 70 & 2 \\
{[1]} & 170 to 180 & 2 \\
{[2]} & 90 to 110 & 37 to 47 \\
{[3]} & 130 to 150 & 37 to 47 \\
{[4] to [7]} & Random inside the polygon & Random inside the polygon \\

\multicolumn{3}{@{}l}{\textbf{\Cref{fig:training_data}e}} \\
{[0]} & 0 & 2 to 16 \\
{[4]} & 240 & 2 to 16 \\
{[1] to [3]} & 0 to 240 & 2 to 16 \\

\bottomrule
\end{tabular}
\label{table:appendix_training_data}
\end{table}

\begin{table}[!htbp]
\centering
\caption{Slope system dimensions of 16 liquefaction-induced flow failure case histories approximated from \cite{olson2001liquefaction_kinetics}, \cite{weber2015engineering_imm}, \cite{macedo2024runout_mpm_cadia}, and \cite{moss2019flow}.}
\label{table:appendix_dam_dimensions}
\begin{tabular}{@{}llll@{}}
\toprule
\multicolumn{1}{c}{Case history name} & \multicolumn{1}{c}{Length (m)} & \multicolumn{1}{c}{Height (m)} & \multicolumn{1}{c}{Informing training data} \\
\midrule
Hachiro-Gata Roadway Embankment & 30 & 3.81 & \Cref{fig:training_data}a \\
La Marquesa Dam & 26 & 8 & \Cref{fig:training_data}c \\
Lake Ackerman Highway Embankment & 42 & 9 & \Cref{fig:training_data}a \\
Chonan Middle School & 33 & 9 & \Cref{fig:training_data}a \\
Route 272 Roadway Embankment & 22.5 & 9 & \Cref{fig:training_data}a \\
Uetsu Line Railway Embankment & 45 & 9.3 & \Cref{fig:training_data}a \\
La Palma Dam & 42 & 12 & \Cref{fig:training_data}c \\
Shibecha-Cho Embankment & 75 & 15 & \Cref{fig:training_data}a \\
Las Palmas Dam & 220 & 15 & \Cref{fig:training_data}b \\
Soviet Tajik Slope & 135 & 29.4 & \Cref{fig:training_data}a \\
North Dike of Wachusett Dam & 120 & 30 & \Cref{fig:training_data}a \\
Takarazuka Landslide & 180 & 30 & \Cref{fig:training_data}a \\
Lower San Fernando Dam & 240 & 53 & \Cref{fig:training_data}c \\
Fort Peck Dam & 450 & 60 & \Cref{fig:training_data}a \\
Calaveras Dam & 385 & 75 & \Cref{fig:training_data}c \\
Cadia Dam & 300 & 80 & \Cref{fig:training_data}b \\
\bottomrule
\end{tabular}
\end{table}

\section{Descriptions of case histories}\label{sec:appendix_case_history}

\subsection*{Lower San Fernando dam}
\begin{figure}
    \centering
    \includegraphics[width=0.5\linewidth]{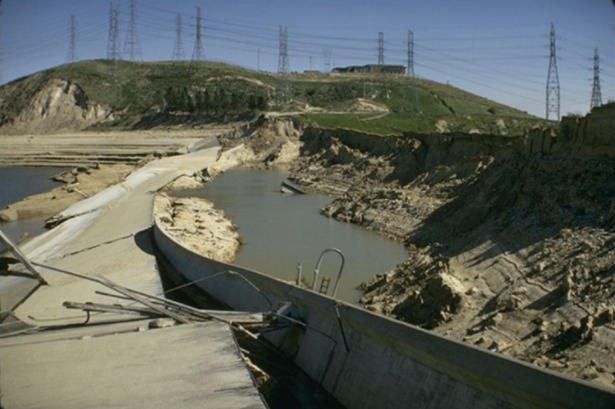}
    \caption{Lower San Fernando dam failure \citep{chowdhury2018evaluation}}
    \label{fig:lsfd_failure_photo}
\end{figure}

The Lower San Fernando dam experienced a catastrophic liquefaction-induced flow failure during the 1971 San Fernando earthquake in California \cref{fig:lsfd_failure_photo}. Previous studies \citep{seed1973analysis_lsfd, seed1975dynamic_analysis_lsfd, olson2001liquefaction_kinetics} indicated that liquefaction in hydraulic fill material of the upstream slope is the main cause of dam failure. Post-earthquake inspections revealed that a substantial portion of the upstream (US) shell and core were mobilized, leaving only 1-2 meters of freeboard. In response, authorities conducted an emergency drawdown of the reservoir and evacuated approximately 80,000 downstream residents. The displaced mass traveled approximately 42–65 meters into the reservoir \citep{talbot2024modeling}, similar to the dam’s original height, and the crest experienced vertical settlements of up to 20 meters, underscoring the severity of deformation. This case has since become one of the most well-studied examples of liquefaction-induced flow failure and serves as a canonical benchmark for evaluating $S_r$ through back-analysis.

\begin{figure}
    \centering
    \includegraphics[width=1.0\linewidth]{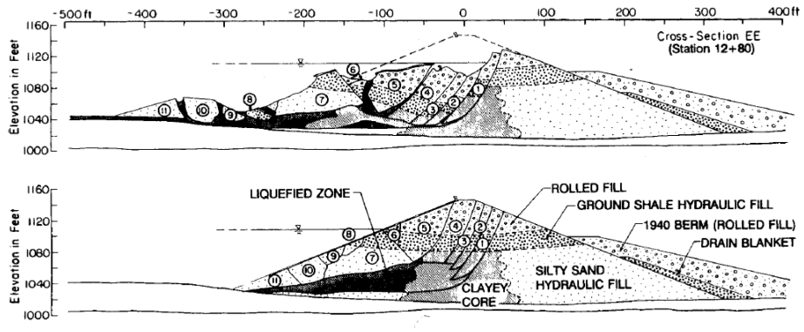}
    \caption{Pre- and post-failure cross-section of the Lower San Fernando dam \citep{seed1989re_lsfd}}
    \label{fig:lsfd_section}
\end{figure}

\Cref{fig:lsfd_section} shows the pre- and post-failure geometry of the dam. It consists of rolled fill, ground shale, clay core, upstream hydraulic fill, downstream hydraulic fill, drain blanket, and berm. The black shaded area in the lower figure indicates the liquefied region of the upstream hydraulic fill.

\subsection*{La Marquesa dam}
\begin{figure}
    \centering
    \includegraphics[width=1.0\linewidth]{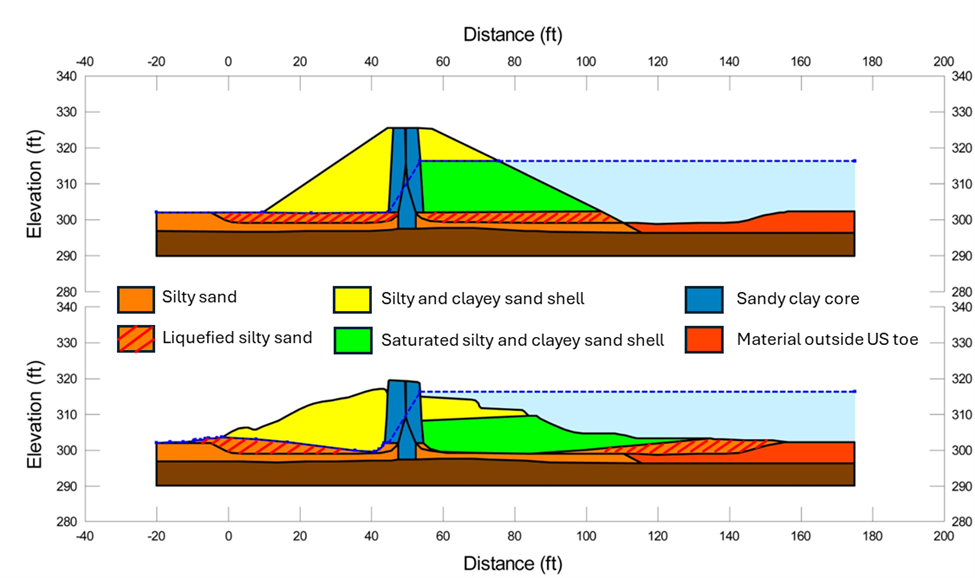}
    \caption{Pre- and post-failure cross-section of the La Marquesa dam (modified after \cite{weber2015engineering_imm})}
    \label{fig:lamqsa_section}
\end{figure}

The La Marquesa dam experienced liquefaction-induced slope failures on both its upstream and downstream sides during the 1985 Central Chilean earthquake \citep{de1988analyses_lamqsa}. Liquefaction within a thin, very loose silty sand layer in the foundation is considered the primary cause of the upstream failure. Displacements were more pronounced on the upstream (US) side, where partial excavation of the foundation toe steepened the slope. Measured displacements included approximately 2 m vertical movement at the crest and 11 m horizontal runout at the toe. The downstream (DS) side experienced relatively smaller, with 6.5 m of the toe runout. \Cref{fig:lamqsa_section} shows the pre- and post-failure geometry. The dam consists of a sandy clay core at the center with shells composed of silty and clayey sands. 

\end{document}